\documentclass[12pt]{iopart}

\usepackage{iopams}
\usepackage{float}
\usepackage{amssymb}
\usepackage{graphicx}
\usepackage{bm}
\usepackage{bbold}
\usepackage[caption = false]{subfig}
\usepackage{xcolor}
\usepackage{cite}

\begin{document}

\title[Inst. of complex fluids with part. structured and part. random interactions]{Instabilities of complex fluids with partially structured and partially random  interactions}

\author{Giorgio Carugno$^{1,2}$, Izaak Neri$^{1,3}$, Pierpaolo Vivo$^{1,4}$}
\address{$^1$ Department of Mathematics, King's College London, Strand, London, WC2R 2LS, UK\\
}
\eads{$^{2}$\mailto{giorgio.carugno@kcl.ac.uk}, \mailto{$^{3}$izaak.neri@kcl.ac.uk}, $^{4}$\mailto{pierpaolo.vivo@kcl.ac.uk}}

\begin{abstract}
We develop a theory for thermodynamic instabilities of complex fluids composed of many interacting chemical species organised in families.   This model includes partially structured and partially random interactions and can be solved exactly using tools from random matrix theory. The model exhibits three kinds of fluid instabilities: one in which the species form a condensate with a local density that depends on their family (family condensation); one in which species demix in two phases depending on their family (family demixing); and one in which species demix in a random manner irrespective of their family (random demixing).   We determine the critical spinodal density of the three types of instabilities and find that the critical spinodal density is finite for both family condensation and family demixing, while for random demixing the critical spinodal density grows as the square root of the number of species.   We use the developed framework to describe phase-separation instability of the cytoplasm induced by a change in  pH.
\end{abstract}

\noindent{\it Keywords\/}: Phase separation, Random Matrix Theory, Block structure, Fluid Instabilities

\maketitle
 
 \section{Introduction}

 Eukaryotic cells are compartmentalised into membrane-bound regions called organelles.  Recently it was found that the cytoplasm also contains  membraneless organelles that form through  liquid-liquid phase separation and are important for various physiological processes \cite{Brangwynne}.\\ \indent
Liquid-liquid phase separation in the cytoplasm is reminiscent of phase separation of  a mixture of oil and water \cite{hymanrev}, but there are also a couple of important distinctions.  Notably, the cytoplasm is composed of a large number of distinct macromolecules \cite{Sear_2005}, e.g.,  human cells contain about $10^9$ proteins from about $10^4$ protein coding genes \cite{proteinsnumber}.   Although  phase separation of two component mixtures, such as oil and water, is well  understood, little is known about the physical principles that govern phase separation of  fluids composed of a large number of distinct molecular species \cite{searcuesta2003, jacobs2013predicting, frenkel1, jacobs2, graf2021thermodynamic}.    The latter problem is also of interest in  other  contexts, such as, for the formation of lipid rafts and clusters of receptors on the cell membrane \cite{membranerafts,lipidrafts, receptorcluster}, the assembly of protein complexes \cite{sartori2020lessons}, the study of polydisperse fluids  \cite{decastro2017,decastro2019,fasolo}, the dynamics of interfaces in crude oil/brine mixtures  \cite{crudeoil}, and the nucleation of iron in the mantle \cite{magma}.  \\ \indent
In an attempt to describe phase separation in a complex fluid, Sear and Cuesta \cite{searcuesta2003} considered a fluid of $N$ components described by a matrix of second order virial coefficients that is random.  Building on random matrix theory \cite{wigner,vivobook},   they found two possible instabilities  of the homogeneous state leading to phase separation of the fluid: one akin to a liquid-vapour coexistence, where the composition of coexisting phases is similar but the total density differs (\emph{condensation}); the other akin to the coexistence of water and oil, where the compositions of the two phases are very different (\emph{demixing}).\\ \indent
Although very successful in predicting possible transitions within a fairly simple and elegant framework, the model in \cite{searcuesta2003} presents some important drawbacks: (i) while condensation happens at a finite critical spinodal density, demixing happens at a total density diverging as $\sqrt{N}$. In other words, the Sear-Cuesta liquid composed of a large number of constituents -- for any practical purpose -- never demixes; (ii) The interaction between species is assumed to be \emph{structureless}, which  neglects important factors playing a role in the affinity/repulsion between molecules observed in reality. For example, (a) the number and geometry of interaction sites in proteins \cite{frenkel2}, and (b) features of the solvent, such as the pH and salt concentration, which affect the fluid particles' net charge.\\ \indent
In order to overcome the above drawbacks, we introduce in this work a model of a complex fluid containing a matrix of virial coefficients that is partially random and partially structured.    This model allows us to embed physically motivated constraints in an otherwise random model, and to derive analytical conditions for the critical spinodal density and the nature of the instability.   As we will show, in a partially random and partially structured  model a complex fluid can demix at finite critical spinodal density, which is consistent with what is observed in experiments in cell biology. \\ \indent
The manuscript is structured as follows. In section \ref{sec:2}, we review the thermodynamics of complex fluids made by many components. In section \ref{sec:3}, we introduce the framework studied in this paper based on matrices of virial coefficients that are partially structured and partially random. In section \ref{sec:4}, we develop a spectral theory to determine the critical spinodal density and the different types of fluid instabilities.  
In section \ref{sec:5}, we illustrate the theory for the specific case of fluids with two families.  In section \ref{sec:6}, we use partially structured and partially random models to describe phase separation of the cytoplasm induced by a change in pH.  In particular, we group proteins in the cytoplasm into an acidic and a basic family according to their response to a change in pH.  We discuss the type of instabilities that can occur and use experimental data available in the literature to give an estimate of the dependence of the critical spinodal density on pH.  Finally, in section \ref{sec:conclusion}, we discuss the assumptions our framework is based on, highlighting both its benefits and its limitations. We conclude by analysing how our methods could be applied to different settings.

\begin{figure} 

    \includegraphics[width = 8cm]{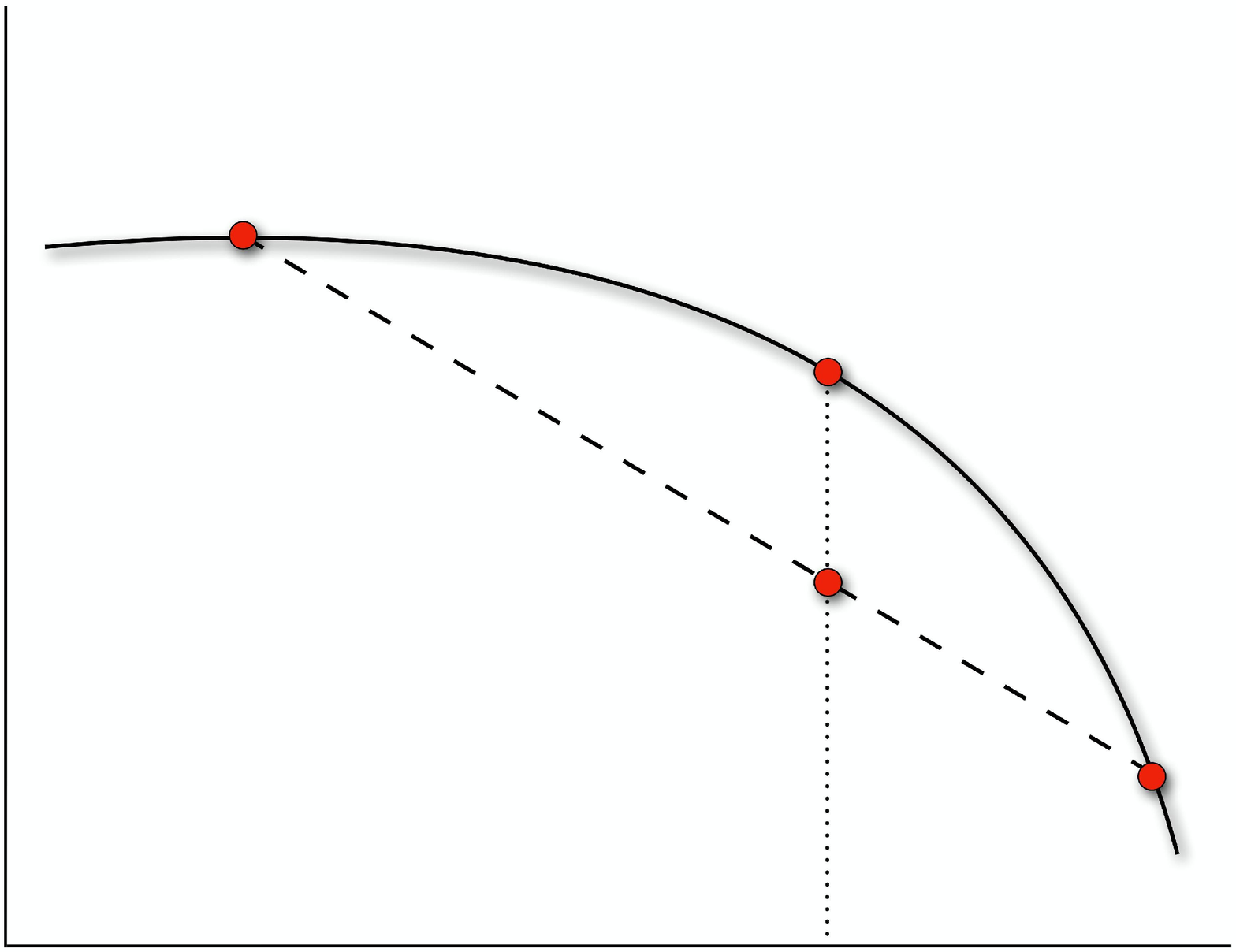}
    \quad 
    \hspace{0.5cm}
    \includegraphics[width = 6cm, height = 6cm]{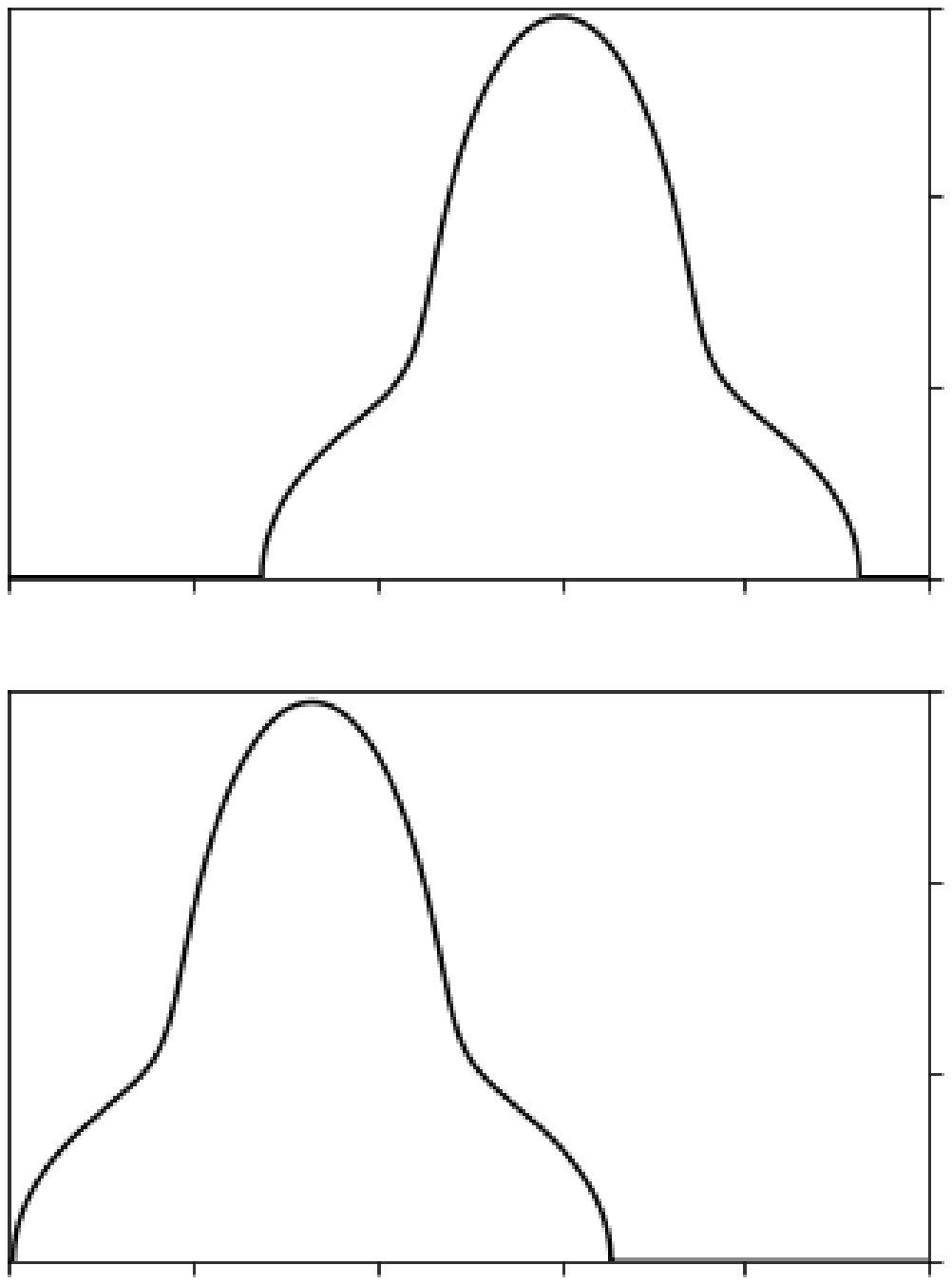} 
    \put(-230,-2){$\rho$}
    \put(-283,-2){$\rho_0$}
    \put(-439,150){ $f$}
    \put(-400,144){$f(\rho_1)$}
    \put(-287,126){$f(\rho_0)$}
    \put(-218,35){$f(\rho_2)$}
    \put(-300,60){$f_{\rm het}$}
    \put(-428.6,169.5){$\wedge$}
    \put(-214,5.08){$>$}
    \put(0,4){$0$}
    \put(-2,25){$0.1$}
    \put(-2,50){$0.2$}
    \put(-2,74){$0.3$}
    \put(0,90){$0$}
    \put(-2,115){$0.1$}
    \put(-2,140){$0.2$}
    \put(-2,164){$0.3$}
    \put(-169,-6){$0$}
    \put(-137,-6){$2$}
    \put(-104,-6){$4$}
    \put(-73,-6){$6$}
    \put(-41,-6){$8$}
    \put(-15,-6){$10$}
    \put(-60,40){$\rho = \rho_c$}
    \put(-153,130){$\rho < \rho_c$}
    \put(-130,108){$\lambda_-$}
    \put(-125,96){$|$}
    \put(-88.5,-10){$\lambda$}
    \put(20,130){$\nu_H(\lambda)$}
    \put(20,40){$\nu_H(\lambda)$}
    
    \caption{ Left: Sketch of the neighbourhood of a state $\rho_0$ in the case of a fluid with one component ($N=1$). The left panel shows the free energy of  homogeneous states (the parabola) and the free energy of  heterogeneous states consisting of two phases with densities $\rho_1$ and $\rho_2$ (the straight line).   The vertical dotted line indicates the corresponding free energies when the total density is $\rho_0$, and in this case $f_{\rm het}<f(\rho_0)$ implying that the free energy is locally unstable.  The right panel sketches  the spectrum of the Hessian of the free energy for $\rho < \rho_c$ (top) and $\rho = \rho_c$ (bottom), where $\rho$ is the total density. For simplicity, we consider a spectrum without outlier eigenvalues described by the spectral density $\nu_H(\lambda)$ -- see \ref{appendix:spectrumblock} for a definition of the spectral density. In the top right panel,  the system is locally stable ($\rho<\rho_c$) implying that the Hessian is positive definite -- $\lambda_- > 0$.  In the lower left panel, the system is marginally stable ($\rho=\rho_c$) and the minimum eigenvalue $\lambda_-=0$.  }
    \label{fig:unstandhess}
\end{figure}

\section{Instabilities in fluids with a large number of components}\label{sec:2}

We consider a fluid mixture at equilibrium composed of $N$ chemical species with number densities $\rho_i$, $i=1,2,\ldots,N$.  In a homogeneous state, i.e.~a state for which all chemical species are well mixed, the free energy density $f$ is in the dilute limit well approximated by a second order virial expansion \cite{mmtheory,mmtheoryextended,landau5,callen1998thermodynamics,Zamponinotes}
\begin{equation}\label{freeenergydensity}
    \beta f(\vec{\rho}) = \sum_{i = 1}^N \rho_i (\log \rho_i -1) + \frac{1}{2} \sum_{i,j=1}^N \rho_i B_{ij} \rho_j \; ,
\end{equation}
where $\beta$ is the inverse temperature (which we  set to $\beta=1$ without loss of generality) and $\vec{\rho}=(\rho_1,\rho_2,\ldots,\rho_N)$. The first term is the equilibrium free energy of an ideal non-interacting gas. The second term, in which matrix elements $B_{ij}$ are the second order virial coefficients, contains information on interactions and characterises the lowest order deviation from ideality of our mixture. The matrix $\mathbf{B}$ is a symmetric matrix with real entries, and has thus real eigenvalues $\{ \lambda_1, \dots, \lambda_N\}$. The spectrum of $\mathbf{B}$ will play a prominent role in the rest of this work. \\ \indent
In principle higher order terms could be added to the expansion, which can be carried out systematically -- \cite{landau5}, see \ref{appendix:virial}.  For simplicity we neglect terms of order higher than two in the virial expansion, but we will come back on  this assumption in Sec.~\ref{sec:conclusion}.\\ \indent
We aim to characterise when a liquid that is homogeneous is unstable towards small spatial fluctuations leading to a heterogeneous state.       In  a heterogeneous state, the fluid is spatially partitioned into $R$ different phases, each of which is homogeneous.   Hence the heterogeneous state is uniquely identified by the densities of chemical species in each phase, $\rho_i^{\alpha}$, and the volume fraction of the different phases $\phi_{\alpha}$, with $\alpha \in \{1, \dots, R\}$ and $\sum_{\alpha} \phi_{\alpha} = 1$.  The  free energy of a heterogeneous state can be expressed as \\ \indent
\begin{equation}
     \beta f_{\rm het}  =  \sum^{R}_{\alpha=1} \phi_{\alpha}  \beta f(\vec{\rho}_{\alpha})\ .
\end{equation} 
A homogeneous state $\vec{\rho}$ can be stable, metastable or unstable: it is stable when all heterogeneous states that can in principle be constructed with total density $\vec{\rho}$ have a higher free energy; it is metastable when the free energy is minimized by a heterogeneous state but there is a finite free energy barrier that the system has to overcome to relax to this heterogeneous state, so that a rare fluctuation is needed to trigger a nucleation event initiating phase separation; it is unstable when the free energy is minimized by a heterogeneous state and there is no free energy barrier keeping the system from relaxing to the minimum -- in this unstable state even an infinitesimal fluctuation would drive the system away from the homogeneous state towards a heterogeneous one.  \\ \indent
The no-free energy barrier condition can be expressed mathematically as follows. Let the fluid be in a homogeneous phase $\vec{\rho}_0$. If there exist near homogeneous states with $\|\vec{\rho}_{\alpha}-\vec{\rho}_{0}\|\ll 1$, for all $\alpha$, such that $f_{\rm het}<f(\vec{\rho_0})$, then we say that the phase $\vec{\rho}_0$ is locally unstable.   This condition materialises when the free energy density is locally not convex -- see Fig.~\ref{fig:unstandhess} left panel. In other words, a phase with  number densities $\vec{\rho}=(\rho_1,\ldots,\rho_N)$ is unstable if the Hessian $H_{ij} = \partial_{\rho_i}\partial_{\rho_j}f$  of the free energy  evaluated at $\vec{\rho}$ has at least one negative eigenvalue. The boundary of the unstable region in density space is called the \textit{spinodal} \cite{Sollich2001} and is determined by the values of $\vec{\rho}$ at which the smallest eigenvalue of $H_{ij}$ is zero.  
For the free energy in (\ref{freeenergydensity}) the Hessian takes the simple form 
\begin{equation}\label{eq:Hessian}
     H_{ij} = \partial_{\rho_i}\partial_{\rho_j} f =  B_{ij} + \delta_{ij} \frac{1}{\rho_i} \ .
\end{equation}  

In what follows, we consider fluid mixtures for which the homogeneous phase, also called parent phase or reference state, has densities $\rho_i$  that are equal to $\rho/N$ with $\rho$ the total density, i.e.,  $\vec\rho_0=(\rho/N,\ldots,\rho/N)$.  In this setup, the only free parameter is the total density $\rho$, which simplifies the analysis of fluid instabilities. Nonetheless, a uniform reference state considerably simplifies the problem, as the spectrum of $\mathbf{H}$ is simply the spectrum of $\mathbf{B}$ shifted by $N/\rho$. Calling $\{\eta_1, \dots, \eta_N\}$ the eigenvalues of $\mathbf{H}$ and $\{\lambda_1, \dots, \lambda_N\}$ the eigenvalues of $\mathbf{B}$, one has

\begin{equation}\label{eq:eigenvaluesHessian}
    \eta_i = \lambda_i + \frac{N}{\rho} \ .
\end{equation}
Let us consider a very dilute phase $\vec{\rho}_0$, that is a phase with small $\rho$. The Hessian of such a phase is dominated by the diagonal term $ \delta_{ij} (1/\rho_i)$ and is thus positive definite. From this phase we increase the total density $\rho$, so that the spectrum of the Hessian starts drifting towards the negative semiaxis -- see Fig.~\ref{fig:unstandhess}, right panel. When the smallest eigenvalue $\eta_-$ of $\mathbf{H}$ touches zero, we have reached the spinodal. This will happen at a critical spinodal density $\rho_c$ given by (see (\ref{eq:eigenvaluesHessian}))

\begin{equation}\label{critdens}
    \rho_c = - \frac{N}{\lambda_-}\ ,
\end{equation} 
where $\lambda_-<0$ is the smallest eigenvalue of $\mathbf{ B}$. We remark that the phase $(\rho_c/N, \dots, \rho_c/N)$ is not a critical point in the sense of second order phase transitions, since one can construct heterogeneous states with a lower free energy -- see fig. \ref{fig:unstandhess}, left panel. \\ \indent
To characterise the nature of a spinodal instability we use the local demixing angle $\theta$ defined as the angle between the reference phase $\vec{\rho_0}$ and the eigenvector $\vec{v}^-$ of the Hessian $\mathbf{H}$ relative to the $0$ eigenvalue, also referred to as the unstable mode

\begin{equation}\label{localangle}
    \theta = \mathrm{min}\{\tilde{\theta}, \pi - \tilde{\theta}\}, \; \tilde{\theta} = \mathrm{arccos}\left(\frac{ \vec{1}\cdot \vec{v}^-}{ \sqrt{N} |\vec{v}^-|}\right),  
\end{equation}
where $\cdot$ is the dot product, $\vec{1}$ is the $N$-dimensional vector of components $[\vec{1}]_i = 1$. The condition $\theta = \mathrm{min}\{\tilde{\theta}, \pi - \tilde{\theta}\}$ ensures that $\theta \in [0,\pi/2]$.  We note that using the uniform reference state $\vec{\rho_0}$, $\vec{v}^-$ coincides with the eigenvector of $\mathbf{B}$ relative to its smallest eigenvalue $\lambda_-$.\\ \indent
The local demixing angle gives information on the type of phase separation when the heterogeneous state is composed by $R=2$ homogeneous phases, so we will focus on this case throughout this work.  \\ \indent
There exist two extreme classes of fluid instabilities \cite{searcuesta2003, frenkel1}: condensation and demixing. In a condensation instability the relative composition is preserved while the total densities of the initial and daughter phases are different: geometrically, the reference phase $\vec{\rho_0}$ and the unstable mode are parallel, that is $\theta = 0$. To give a physical intuition, an example of a condensation instability would be the transition from vapour to liquid water after a drop in temperature. In a demixing instability, the two daughter phases have fully distinct compositions but the same total number density: geometrically,  the reference phase $\vec{\rho_0}$ and the unstable mode are perpendicular, that is $\theta = \pi/2$. A physical example would be a mixture of oil and water in equal parts, which is initially made homogeneous by stirring it thoroughly: immediately after one stops stirring, this homogeneous state becomes unstable and droplets of water and oil will form. Intermediate situations between condensation and demixing are possible, and are characterised by a value of $\theta\in (0,\pi/2)$. \\ \indent
The local demixing angle $\theta$ -- strictly speaking only meaningful \emph{at the onset} of a spinodal transition -- is
effective in predicting the type of instability at equilibrium \cite{frenkel1}, even though it can be quantitatively different from the ``final" angle \cite{fasolo}. Given a sensible model for the interaction matrix $\mathbf B$, spinodal instabilities depend therefore only on its smallest eigenvalue and associated eigenvector. \\ \indent

\section{Partially structured and random virial coefficients} \label{sec:3}

Computing virial coefficients from first principles for a mixture composed by a large number of interacting species is not feasible, and experimental data are scarce. For this reason, in line with a long tradition of modelling complex systems with random matrices \cite{may, wigner, bouchaudfinance, searcuesta2003}, we adopt a statistical description and replace the unknown virial coefficients  between the chemical species with random interactions. This can be justified by a ``central limit theorem'' type of argument in the case of biological macromolecules: since biochemical interactions result from the sum of a large number of microscopic terms, assuming they are independent they should be well-approximated by a mean-value effective interaction plus Gaussian fluctuations. This in turn would make the virial coefficients $B_{ij}$ random variables. The main advantage of this approach is that it gives clear theoretical predictions, which can in principle be tested with data. \\ \indent
From a modelling perspective it is desirable to include generic properties of the interactions between the chemical species in the model.   To this aim, we assume that the components of the fluid can be  grouped into a small number of $F$ families according to their  characteristics, such as their charge or their hydrophobicity, and we assume that the statistical properties of the intra- and inter-family interactions are known.  In other words, we assume that the  properties of the virial matrix on the coarse-grained level of families are known, while the detailed microscopic interactions are unknown.  \\ \indent
Mathematically, we describe a partially structured and random model with a virial matrix  of the form
\begin{equation}\label{brepresentation}
    \mathbf{B} =  \mathbf{D} + \mathbf{C}* \mathbf{Z}\ , 
\end{equation}
where $\mathbf{D}$ and $\mathbf{C}$ are deterministic rank-$F$ matrices representing the coarse-grained knowledge we have about the interactions between families, and $\mathbf{Z}$ is a random matrix  that represents our ignorance about the microscopic interactions.  The symbol $*$ is the element-wise product. This decomposition is particularly useful to study the spectrum of $\mathbf{B}$, as one can link properties of its spectrum to the matrix elements of $\mathbf{D}$ and $\mathbf{C}$. \\ \indent
We label the indices such that the matrices $\mathbf{D}$ and $\mathbf{C}$ have a block structure, in particular, $D_{ij} = \mu_{s(i)s(j)}$ and $C_{ij} = \sigma_{s(i)s(j)}$,  where $s(i)\in \left\{1,2,\ldots, F\right\}$ is a function that keeps track of which family $s$ a given index $i$ belongs to.  Without loss of generality, we can order the indices $i$ such  that $s(i)=s$ for all  $i \in \left\{\sum_{t=0}^{s-1} N_{t}+1,\ldots,\sum_{t=0}^{s} N_{t}\right\}$ where $N_s$ denotes the number of species that belong to family $s$ and $N_0=0$. \\ \indent
The entries of $\mathbf{Z}$ are independent and identically distributed  random variables with mean $0$ and variance $1$.
For $F=1$ we recover the model of Sear and Cuesta \cite{searcuesta2003}.   Partially random and partially structured random matrices have been studied before in the context of neural networks \cite{ahmadian2015properties,lowdimensionalstructuredneural}, but those models are nonsymmetric, whereas in the present case the matrices are symmetric.

\section{Spectral properties of large random matrices with a block structure}\label{sec:4}

The spectrum of an infinitely large matrix of the form (\ref{brepresentation}) consists of two parts, a continuous spectrum determined by the random matrix $\mathbf{C}* \mathbf{Z}$   and a finite number of at most $F$ outlier eigenvalues $\lambda_{\rm isol}$ that are determined by the deterministic matrix $\mathbf{D}$ -- see Fig.~\ref{fig:spectra}.  Both components of the spectrum are deterministic in the limit of large $N$, and hence we do not need to worry about sample-to-sample fluctuations.
Depending on the scaling with $N$ of the moments of $B_{ij}$, outliers can be influenced by the noise. The smallest eigenvalue $\lambda_-$ is either located at the lower edge of the continuous spectrum or is one of the outliers.   The associated eigenvector $\vec{v}^-$, and thus the nature of the instability, has different properties in the two cases.  \\ \indent
In the following we will discuss the fluid instabilities in three versions of the  model (\ref{brepresentation}) with increasing detail in the matrix $\mathbf{C}$ of noise amplitudes: (i) the deterministic  case with zero noise amplitudes; (ii) the  case with uniform noise amplitudes, $\mathbf{C} = \sigma\mathbb{1} $, where $\mathbb{1}$ is the identity matrix; (iii) the general case.   \\ \indent

\subsection{Deterministic case: family condensation and family demixing}\label{subsec:deterministiccase}
\begin{figure}
\centering
 \includegraphics[width = 12cm]{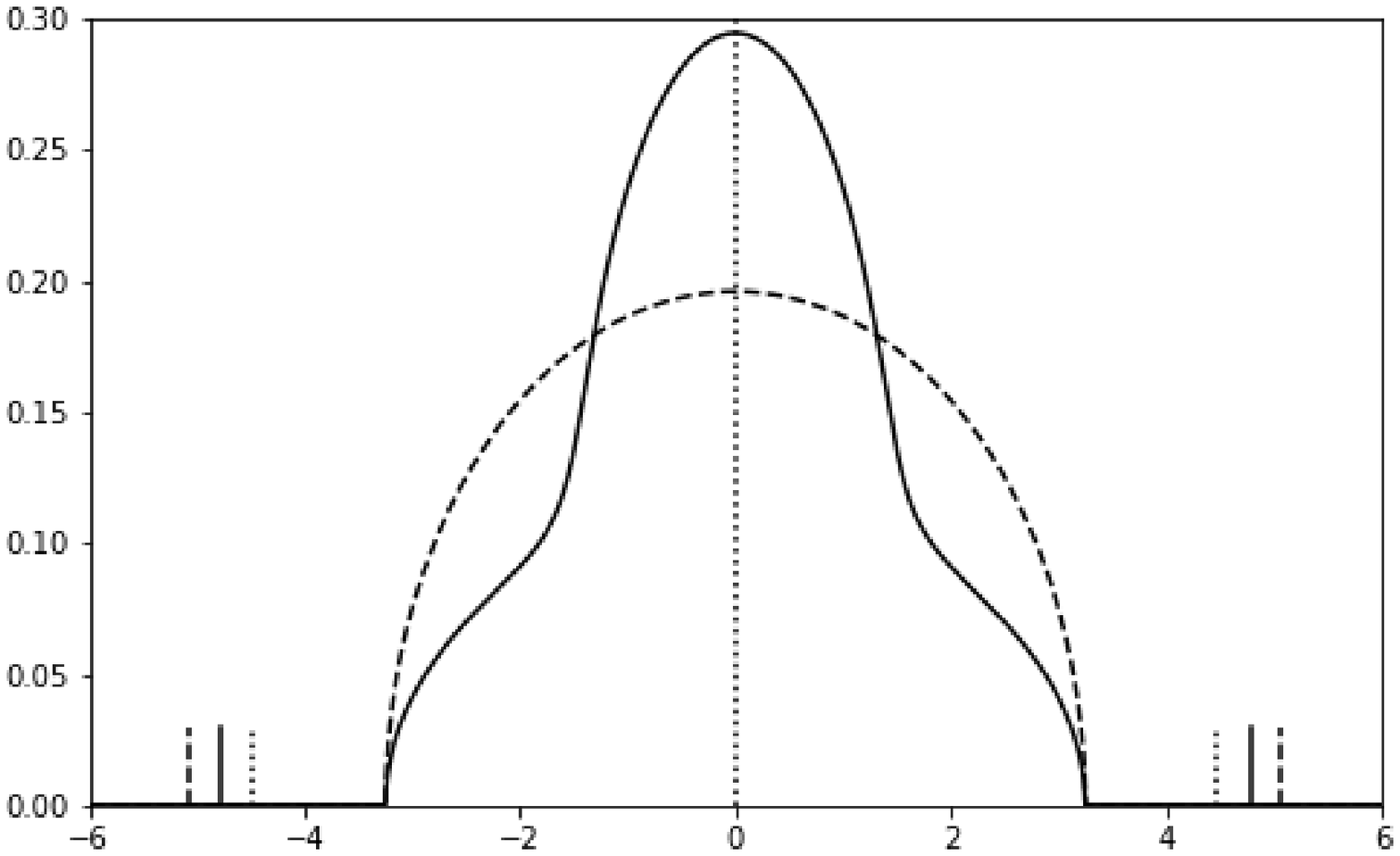}
    \put(-165,-4){$\lambda/N$}
     \put(-360,110){$\nu(\lambda)$}
    \put(-50.74,31.13){$\wedge$}
    \put(-42.5,31.1){$\wedge$}
    \put(-35.5,31.1){$\wedge$}
    \put(-165.7,197.4){$\wedge$}
    \put(-296.3,31.1){$\wedge$}
    \put(-288.9,31.1){$\wedge$}
    \put(-281.41,31.1){$\wedge$}
    
    \caption{Comparison between the spectra of large matrices $\mathbf{B}$ of the form (\ref{brepresentation}) in the case $F=2$  without disorder (dotted line), with  disorder that has uniform (i.e., family-independent) variances (dashed line),  with disorder that has non-uniform (i.e., family dependent) variances (continuous line); for analytical calculation of spectra see \ref{appendix:spectrumblock}. Upward pointing arrows stand for Dirac delta functions: one dotted arrow at $0$ represents $N-2$ degenerate eigenvalues $\lambda = 0$ in the deterministic case; $3$ pairs of deltas of weight $1/N$ represent isolated eigenvalues separated from the bulk. The mean values of the entries of $\mathbf{B}$ are $\mu_{11} = -4.0$, $\mu_{12} = 8.0$, $\mu_{22} = 4.0$, $\mu_{11} = -4.0$, and the fraction of species in the first family is $c_1 = 0.5$ in all cases. The variances of the entries in the non-uniform case are $\sigma^2_{11} = N$, $\sigma^2_{12} = N/2$, $\sigma^2_{11} = 5N$, while in the uniform case $\sigma^2$ is chosen to have the same edge $\lambda_{\mathrm{b}}$ as in the non-uniform case. See \ref{sec:4} and \ref{appendix:spectrumblock} for mathematical details. }
    \label{fig:spectra}
\end{figure}

In the  deterministic case $\mathbf{C} = \mathbf{0}$,  we recover an effective model describing a fluid of $F$ components, corresponding to the $F$ families, and with a virial matrix $\mathbf{B}_{\rm det}$ with entries $[\mathbf{B}_{\rm det}]_{st} = c_t\mu_{st}$, where $c_t$ is the fraction of species belonging to each family and $s,t\in\left\{1,2,\ldots F,\right\}$.   In the present case the spectrum of the matrix $\mathbf{B}$ consists of two parts (see Fig.~\ref{fig:spectra}): (i) a zero eigenvalue with multiplicity equal to $N-F$; and (ii) $F$ nonzero eigenvalues $\lambda_{\rm isol} = \gamma_{\rm isol} N$, where $\gamma_{\rm isol}$ are the eigenvalues of $\mathbf{B}_{\rm det}$.  Also the eigenvectors of   $\mathbf{B}$ are determined by $\mathbf{B}_{\rm det}$: the $F$ eigenvectors associated with  the $F$ isolated eigenvalues $\lambda_{\rm isol}$ have components $v_i$ that depend only on the family to which the species $i$ belongs, i.e., $v_i = V_{s(i)}$, where the  $V_{s}$ are the eigenvector components of $\mathbf{B}_{\rm det}$ associated with the eigenvalues $\gamma_{\rm isol}$.    The  $V_s$ solve the equation
\begin{equation}\label{systemdeterministc}
    V_s = \frac{1}{\gamma_{\rm isol}}\sum_{t=1}^F c_t \mu_{st}V_t \; .
\end{equation} 

Let us now discuss the implications of the spectral properties of $\mathbf{B}$ for phase separation.  
The smallest eigenvalue of $\mathbf{B}$ is either $0$ or the minimum of the $F$ non trivial eigenvalues. In the former case, the critical spinodaldensity (\ref{critdens}) is infinite and therefore the homogeneous phase $\vec{\rho}_0$ is always stable. In the latter case, $\rho_c$ is of order $\mathcal{O}(1)$ for large $N$ as  $\lambda_{\rm isol}$ scales linearly with $N$. \\ \indent
The nature of the liquid instability is determined by the $V_1, \dots, V_F$ that solve (\ref{systemdeterministc}). For $F=1$ the only possible unstable mode is parallel to $\vec{\rho}_0$  and we recover the condensation instability of \cite{searcuesta2003}. For a generic $F$, the possibilities are much richer and it is useful to distinguish two cases. When all $V_s$ have the same sign, then one of the daughter phases is enriched in all species, albeit in proportions depending on their family. We refer to this kind of instability as \emph{family condensation} and $\theta$ is geometrically bounded $0 \leq \theta \leq \mathrm{arccos}(\sqrt{c_{\mathrm{min}}})$, where $c_{\mathrm{min}} = \mathrm{min}_{s} c_s$. Conversely, when not all $V_s$ have the same sign, one daughter phase is enriched in some species and deprived in others, depending on which family they belong to. We refer to this kind of instabilities as \emph{family demixing} and $\theta$ is geometrically bounded $\mathrm{arccos}(\sqrt{c_{\mathrm{max}}}) \leq \theta \leq \pi/2 $, where $c_{\mathrm{max}} = \mathrm{max}_{s} c_s$. In the latter case, there exists a region in the space of parameters for which $\theta = \pi/2$, corresponding to a demixing instability at finite critical spinodal density.  \\ \indent
In the next sections we show that this picture is robust against the addition of disorder to the virial matrix.\\ \indent

\subsection{Uniform noise amplitude}\label{subsec:uniformnoise}

We now add uniform noise (i.e. family-independent) to the matrix of virial coefficients by setting $\mathbf{C} = \sigma \mathbb{1} $.  In this case we can characterise the spectrum of $\mathbf{B}$ analytically as it  corresponds to a finite rank perturbation of a Wigner random matrix $\mathbf{Z}$~\cite{capitaine2009largest}.\\ \indent
As illustrated in Fig.~\ref{fig:spectra}, the main effect of the noise is that the zero eigenvalue transforms into a continuous spectrum  described by the Wigner's semicircle law supported on the interval $[-2\sigma\sqrt{N},2\sigma\sqrt{N}]$~\cite{bouchaud,vivobook,mehta,bai2008methodologies}.   On the other hand, the spectrum of $\mathbf{B}$ can retain some signature of the non-zero eigenvalues of the deterministic matrix $\mathbf{D}$ in the form of eigenvalues isolated from the bulk.  At finite but large $N$ the isolated eigenvalues  are located at  $\lambda_{\rm isol} = N \gamma_{\rm isol} + \sigma^2/(N \gamma_{\rm isol})$, where $\gamma_{\rm isol}$ are as before the eigenvalues of $\bf{B}_{\rm det}$.     Consequently, $\lambda_-$ is either the lower edge of the spectral density $\lambda_{\rm b} = -2\sigma \sqrt{N}$ or the lowest among the  outliers $\lambda_{\rm isol}$. Substituting in (\ref{critdens}) these values for $\lambda_-$ one obtains for the critical spinodal density 
\begin{equation}\label{wigner_critdens}
    \rho_c = \mathrm{min} \left \{\frac{\sqrt{N}}{2\sigma}, -\frac{ \gamma^-_{\rm isol}}{(\gamma^-_{\rm isol})^2 + \sigma^2/N} \right \} \ .
\end{equation} 

Let us now discuss the nature of the instability corresponding to the two cases described in Eq.~(\ref{wigner_critdens}).  When $\lambda_-$ is the lower edge $\lambda_{\rm b}$ of the continuous spectral density, then $\rho_c$ scales as $\sqrt{N}$ and the associated eigenvector is a random vector with Gaussian components. Expression (\ref{localangle}) implies that such an eigenvector is orthogonal to the reference phase $\vec{\rho}_0$, and thus describes a demixing phase transition referred to as {\it random demixing } \cite{searcuesta2003}. On the other hand, when $\lambda_-$ is the smallest outlier $\lambda_{\rm isol}$,  then its associated eigenvector  converges for large $N$ to the corresponding eigenvector in the deterministic model discussed in Sec.~\ref{subsec:deterministiccase}  and thus describes  either  family condensation or family demixing.    Equations (\ref{wigner_critdens}) implies that at finite values of $N$, the crossover between random demixing and family condensation (or family demixing) happens at $\sigma^\ast \approx \sqrt{N}\gamma^{-}_{\rm isol}$, and hence at large values of $N$ the dispersity in the virial coefficients has to be large in order to observe a random demixing transition.   \\ \indent

\subsection{General case}\label{subsec:generalnoise}

In the general case where noise amplitudes are not uniform, the picture is qualitatively equivalent to that of the perturbed Wigner case.   If $\lambda_-$ is located at the edge of the continuous spectrum, then the fluid is unstable w.r.t.~random demixing above a critical spinodal density that grows with the number of different protein species as $\rho_c \sim \mathcal{O}(\sqrt{N})$.  On the other hand,  if $\lambda_-$ is an outlier, then the fluid is unstable w.r.t. family condensation or family demixing above a finite critical spinodal density $\rho_c \sim \mathcal{O}(1)$.  However,   the critical spinodal density $\rho_c$ and the modes $\vec{v}^-$ of instability are, in the case of noise variances that are family-dependent, quantitatively different from  when noise variances are family-independent. \\ \indent
Let us first discuss the cases when $\lambda_-$ is an outlier $\lambda_{\rm isol}$, corresponding to family condensation or family demixing.   The   entries $v_i = V_{s(i)}$ of eigenvectors associated with outlier eigenvalues  solve  \cite{kabacavity, neri2016eigenvalue, susca, izaak,lowrankpert} (see also  \ref{appendix:spectrumblock})
\begin{equation}\label{systemeigenvectors}
    V_s =  G_{s}\left(\frac{\lambda_{{\rm isol}}}{N}\right)  \sum_{t = 1}^F  c_t \mu_{st} V_t, \quad {\rm with} \quad s \in \{1,\ldots, F\} \;  , 
\end{equation}
where the $G_s(\lambda)$  solve \cite{torben}  
\begin{equation}\label{systemresolvent}
    \frac{1}{G_s} = \lambda - \sum_{t = 1}^F c_t \sigma_{st}^2 G_{t}, \quad {\rm with} \quad s \in \{1,\ldots, F\} \; .
\end{equation} 
The eigenvalue outliers are found as the nontrivial solutions (i.e., $\vec{v}^-\neq 0$) of the set of Eqs.~(\ref{systemeigenvectors}-\ref{systemresolvent}), and  $\lambda_-$ is the smallest of those.  For family-dependent noise variances the $G_s$ are dependent on $s$, and therefore the modes $\vec{v}^-$ are in general different from those for family-independent noise variances.\\ \indent
In \ref{appendix:spectrumblock} we  derive Eqs.~(\ref{systemeigenvectors}) and (\ref{systemresolvent}) using the resolvent, defined as the matrix inverse $G_B = (z \mathbb{1} - \mathbf{B})^{-1}$ for $z \in \mathbb{C}/\{\lambda_1, \dots, \lambda_N\}$ \cite{izaak}.  We show that when the size $N$ of the matrix $\mathbf{B}$ diverges, the diagonal elements of the resolvent $[G_{\tilde{B}}(\lambda - \mathrm{i} \epsilon_N)]_{ii}$ -- where $\tilde{\mathbf{B}}$ is a scaled version of $\mathbf{B}$ and  $\epsilon_N$ is a vanishing regulariser, see \ref{appendix:spectrumblock} for details -- tend to the values  $G_{s(i)}(\lambda)$ that solve  the set of Eqs.~(\ref{systemresolvent}) for $z = \lambda$, where $s(i)$ indicates the family to which species $i$ belongs.  We note that while Eqs.~(\ref{systemeigenvectors}) are valid only for an outlier eigenvalue, Eqs.~(\ref{systemresolvent}) are  general.\\ 
Let us now discuss the continuous part of the spectrum of $\mathbf{B}$.  There are two ways to obtain the edge $\lambda_{\rm b}$ of the continuous spectrum.    A first approach determines the  spectral density  $\nu(\lambda)$
\begin{equation}
    \nu(\lambda) = \lim_{N \to \infty} \frac{1}{N} \sum_{i=1}^N \delta(\lambda_i - \lambda) \ ,
\end{equation}
from the diagonal components of the resolvent  of $\mathbf{B}$.  In particular, the normalized trace of the resolvent is the Stieltjes transform of $\nu(\lambda)$ \cite{mehta}, which can be inverted to give
\begin{equation}
    \nu(\lambda) = \lim_{N \to \infty} \lim_{\epsilon \to 0} \mathrm{Im} \left[\frac{1}{N} \mathrm{Tr}\;  G_B(\lambda - \mathrm{i} \epsilon)\right] \ .
\end{equation}

As shown Fig.~\ref{fig:spectra}, for family-dependent amplitudes the continuous part of the spectrum is not a Wigner semicircle, contrarily to the case with family-independent noise amplitudes. \\ \indent
A second approach to determine $\lambda_{\rm b}$ is to study the eigenvectors of $\mathbf{B}$. The  eigenvectors associated with    the edge of the continuous spectrum are random vectors with entries $v_i$  that are drawn independently from  Gaussian distributions with  zero mean and with variances $\Delta_{s(i)}$ that depend on the family $s(i)$ to which the $i$-th index belongs to.  As shown in \ref{appoff}, the variances $\Delta_s$ solve the equations 
\begin{equation}\label{eq:systemvarianceseigvec}
    \Delta_s = G_s^2\left(\frac{\lambda_{\rm b}}{\sqrt{N}}\right) \sum\limits_{t=1}^F c_t \sigma^2_{st} \Delta_t \quad {\rm with} \quad s \in \{1,\ldots, F\} \; .\end{equation}
In the case of uniform noise amplitudes $\sigma_{st} = \sigma$, and we obtain that $\Delta_s=\Delta$. Using expressions available in the literature for the resolvent of a Wigner matrix \cite{mehta}, equation (\ref{eq:systemvarianceseigvec}) has non-trivial solutions if $\lambda = \pm 2 \sigma \sqrt{N}$, in agreement with results from subsection \ref{subsec:uniformnoise}.   
  \\ \indent

\section{Spinodals for two families ($F=2$)}\label{sec:5}

\begin{figure}
    \includegraphics[width = 7cm]{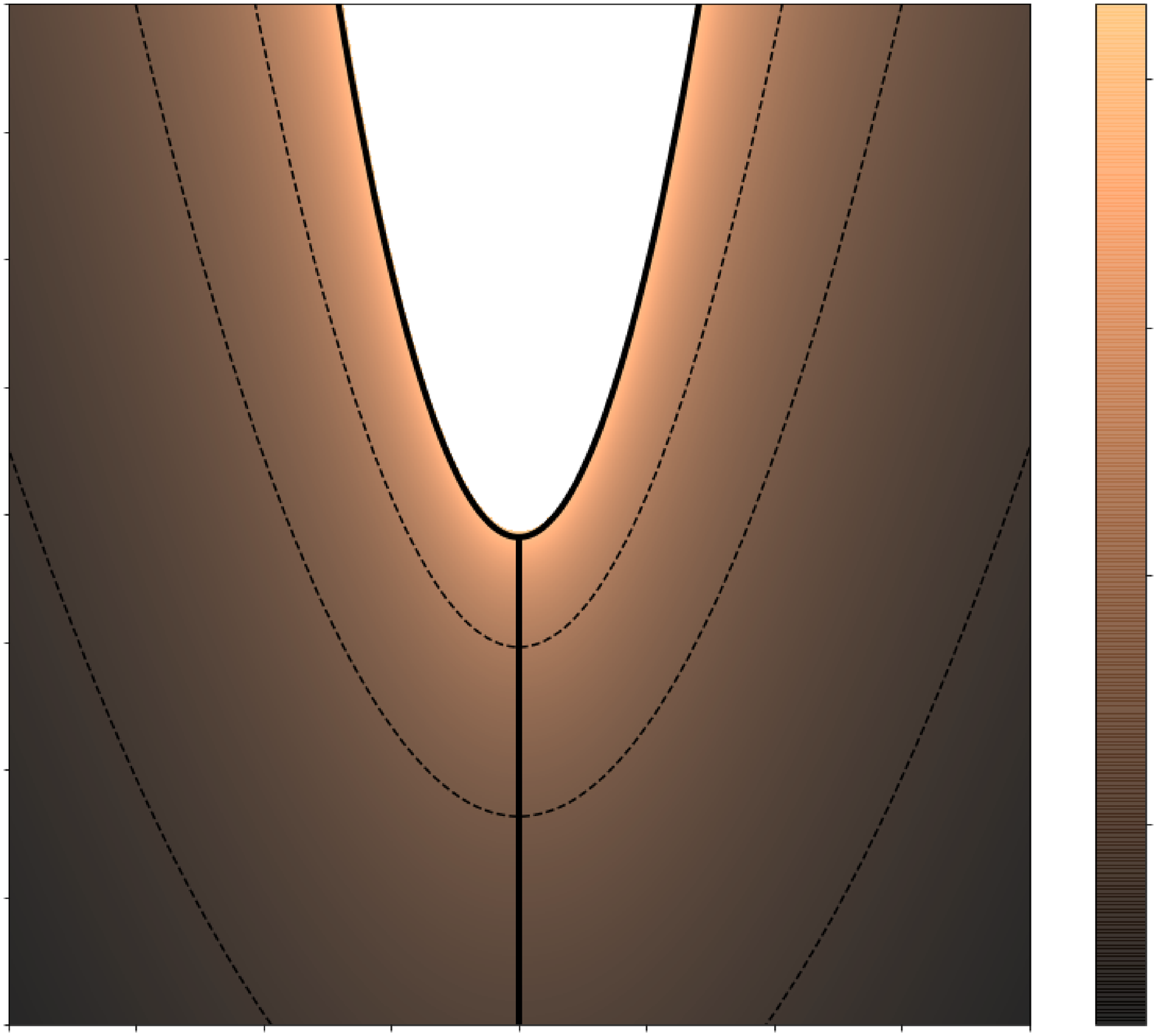}
    \quad 
    \hspace{1cm}
    \includegraphics[width = 7cm]{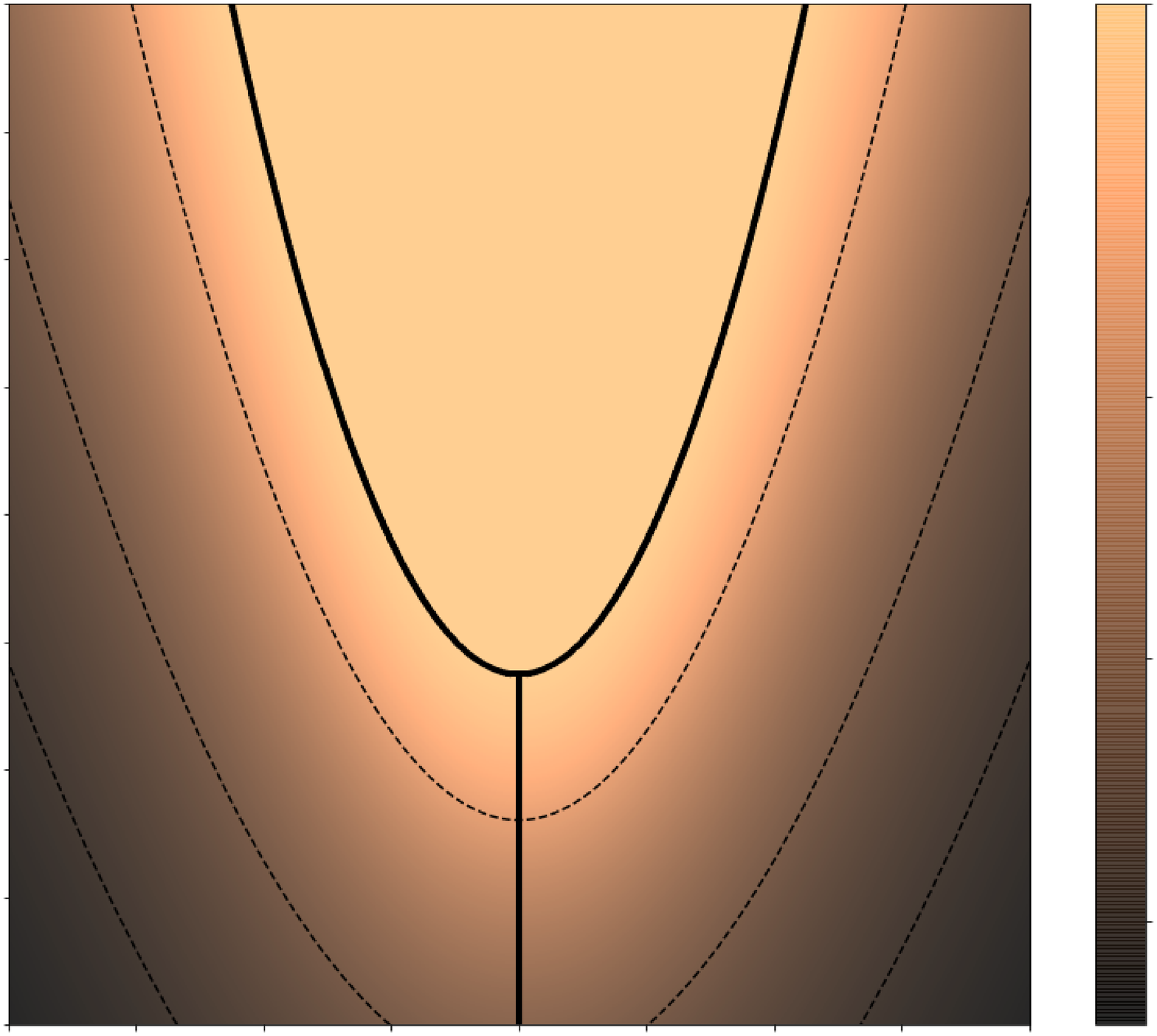} 
     \put(-115,-16){$\mu_{12}$}
     \put(-10,184){$\rho_c$}
     \put(-360,-16){$\mu_{12}$}
     \put(-265,184){$\log\rho_c$} 
     \put(-280,-8){10}
     \put(-358,-8){0}
      \put(-448,-8){-10}
     \put(-111,-8){0}
     \put(-30,-8){10}
      \put(-205,-8){-10}
       \put(-457,170){10}
     \put(-472,85){$\mu_{22}$}
     \put(-452,85){0}
      \put(-462,1){-10}
      \put(-412,60){\textcolor{white}{\textbf{FC}}}
      \put(-315,60){\textcolor{white}{\textbf{FD}}}
      \put(-160,60){\textbf{FC}}
      \put(-75,60){\textbf{FD}}
      \put(-120,130){\textbf{RD}}
       \put(-359,130){\textbf{S}}
       \put(1,173){0.55}
       \put(1,63){0.3}
       \put(-246,161){2}
       \put(-246,76){0}
      
    \caption{ Comparison between the spinodals of complex fluids with deterministic interactions (left) and those with random interactions (right).    The thick line separates different types of instability, in particular, family condensation (\textbf{FC}), family demixing (\textbf{FD}), random demixing (\textbf{RD}), and a region of stability (\textbf{S}) in white, and the thin lines denote contours of constant $\rho_c$. In the left picture, the color bar is in logarithmic scale.  Parameters are $F=2$, $\mu_{11} = 1$ and $c_1 = 0.7$,   $\sigma_{11} = \sigma_{21}=\sigma_{22} = 0$ (left),  $\sigma_{11}^2 = N$, $\sigma_{12}^2 = 0.5 N$, and $\sigma_{22}^2 = 1.5 N$ (right), where $N$ is the number of species.  }
    \label{fig:deterministic}
\end{figure}

We use the  case of two families $F=2$ to illustrate the influence of randomness on fluid instabilities.  \\ \indent
Figure \ref{fig:deterministic} shows the critical spinodal density $\rho_c$ above which  the homogeneous state is unstable,
and the figure also indicates the nature of the instability, i.e., whether the instability of the homogeneous state is towards a heterogeneous state with family condensation, family demixing, or random demixing.   The left panel considers the case of a deterministic matrix  of virial coefficients, whereas the right panel shows what happens in the case that the matrix of virial coefficients is random (we consider the general case of nonuniform noise amplitudes).   Comparing the deterministic case (left panel) with the random case (right panel), we observe that in the deterministic case there exists a region that is stable at all densities, whereas in the random case this region is destabilised by the random demixing phase.\\ \indent\\ \indent
Note that in order to have a finite critical spinodal density towards demixing at large values of $N$, we have scaled the  variances of the virial coefficients  linearly with $N$.  Otherwise, the critical spinodal density towards random demixing diverges as $\sqrt{N}$, see Eq.~(\ref{wigner_critdens}).

\begin{figure}
    \includegraphics[width = 6.5cm]{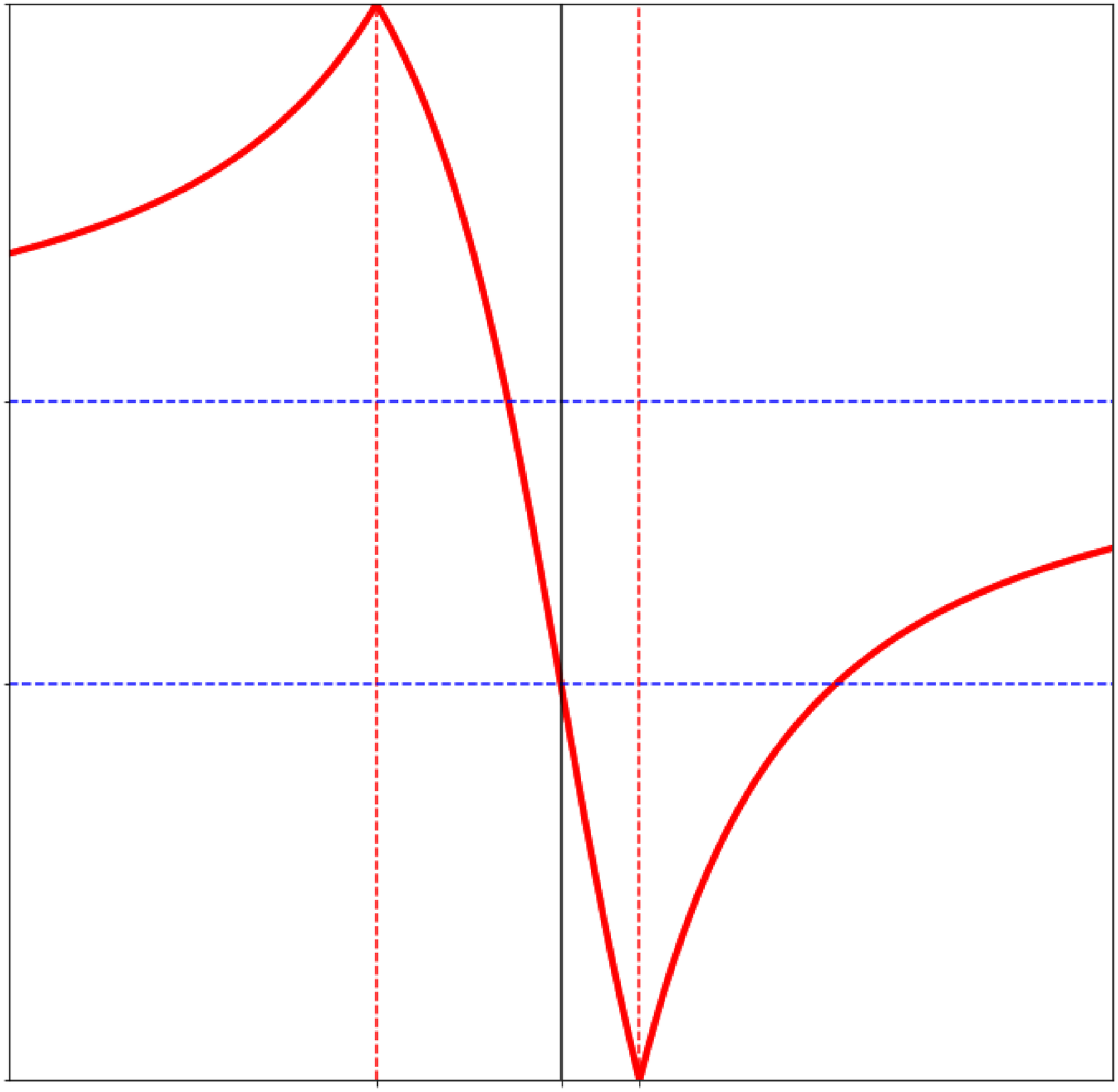}
    \quad
    \hspace{1cm}
    \includegraphics[width = 7.2cm]{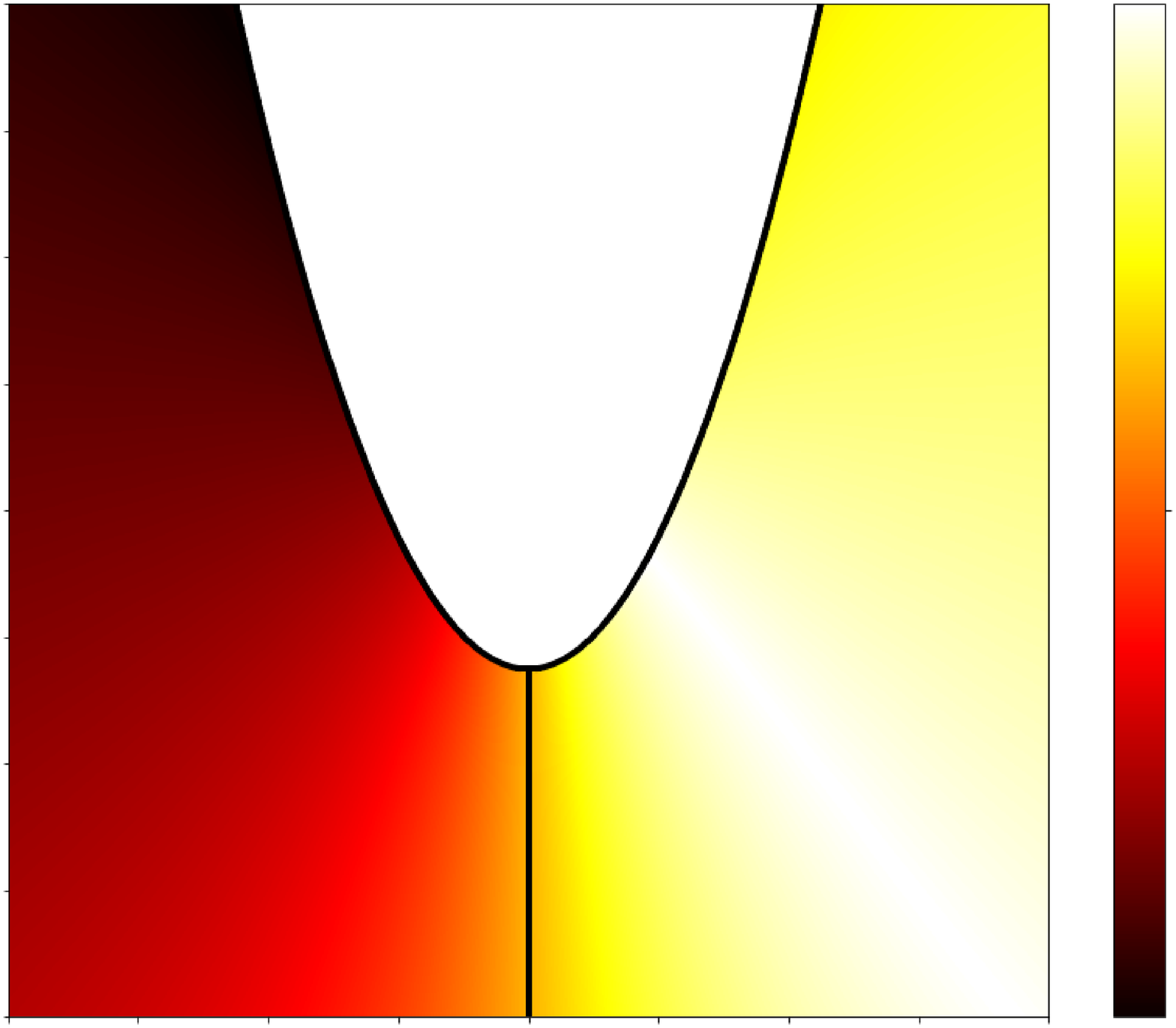}
    \put(-348,-19){$\zeta$}
      \put(-231,86){$\mu_{22}$}
     \put(-120,-17){$\mu_{12}$}
     \put(-451,86){$\theta$} 
     \put(-444,1){0}
      \put(-444,175){$\frac{\pi}{2}$}
     \put(-30,-8){10}
     \put(-115,-8){0}
      \put(-210,-8){-10} 
       \put(-217,170){10}
     \put(-211,85){0}
      \put(-222,2){-10}
      \put(-165,60){\textcolor{white}{\textbf{FC}}}
      \put(-70,60){\textbf{FD}}
      \put(-122,130){\textbf{RD}}
      \put(-348,-8){0}
     \put(-335,-8){1}
      \put(-390,-9){$-\frac{c_2}{c_1}$} 
      \put(-446,63){$\theta_2$}
      \put(-446,110){$\theta_1$}
      \put(1,1){0}
      \put(1,175){$\frac{\pi}{2}$}
       \put(-9,182){$\theta$}
    \caption{Demixing angle $\theta$ characterising the instability of a complex fluid.  Parameters are set as in Fig.~\ref{fig:deterministic}. On the left, $\theta$ as a function of $\zeta$ for $c_1 = 0.3$. Extremes (red) and asymptotes (blue) are highlighted by dashed lines. On the right, a heathmap of $\theta$: in the \textbf{FC} region, $\zeta > 0$ and  $\theta < \theta_1 =  \mathrm{arccos}(\sqrt{c_1})$; in the \textbf{FD} region $\zeta < 0$ and  $\theta > \theta_2  =  \mathrm{arccos}(\sqrt{c_2})$; in the \textbf{RD} region $\theta = \pi/2$ (see the caption of Fig.~\ref{fig:deterministic} for the definition of the acronyms).}
    \label{fig:ratio}
\end{figure}

To determine the nature of the  mode that  destabilises the homogeneous state we need to analyse the eigenvector $v^-$ associated with $\lambda_-$. 
Provided that $\lambda_-$ is an outlier, the unstable mode can be cast in the form $(\zeta, \dots, \zeta, 1, \dots, 1)$  with $\zeta = V_1 / V_2$ (see  \ref{appendix:f2spinodal}). Substituting this expression for the unstable mode into (\ref{localangle}), the local demixing angle can be expressed as

\begin{equation}\label{thetazeta}
    \theta = \mathrm{min}\{\tilde{\theta},\pi - \tilde{\theta}\}, \quad {\rm where} \quad  \tilde{\theta} = \mathrm{arccos} \left(\frac{c_1 \zeta + c_2}{\sqrt{c_1 \zeta^2 + c_2}}\right) \ .
\end{equation}
In the left panel of Fig.~\ref{fig:ratio} we plot the local demixing angle $\theta$ as a function of $\zeta$.  This plot shows that the range of values that $\theta$ can take is strongly dependent on the sign of $\zeta$, in agreement with the bounds discussed in Sec.~\ref{subsec:deterministiccase} for family condensation and family demixing.\\ \indent
The dependence on the statistics of $\mathbf{B}$ enters in the expression (\ref{thetazeta}) only through the value of $\zeta$.     In \ref{appendix:f2spinodal} we use the Eqs.~(\ref{systemeigenvectors}-\ref{systemresolvent}) to obtain $\zeta$ as a function of the model parameters.  Substituting $\zeta$ in  Eq.~(\ref{thetazeta}) we obtain the local demixing angle $\theta$, which for the case of  Fig.~\ref{fig:deterministic}  is shown in the right panel of  Fig.~\ref{fig:ratio}.  \\ \indent
Interestingly, in all matrix models -- deterministic, uniform noise and general case -- the only parameter influencing the sign of $\zeta$ is the interfamily average virial coefficient $\mu_{12}$ as $\mathrm{sgn}(\zeta) = - \mathrm{sgn}(\mu_{12})$ (see \ref{app:c3} for more detail). The physical picture is the following: when  $\mu_{12}<0$, then the  interactions between two proteins belonging to different families are  dominantly attractive and therefore there is a net free energy reduction if species of family $1$ and $2$ aggregate together; conversely, when  $\mu_{12}>0$ then the interactions between proteins of different families are dominantly repulsive and therefore  there is a net free energy reduction if the $2$ families demix. Furthermore, we notice a symmetry between family condensation and family demixing in the presented models -- if a system is unstable towards family condensation at a critical spinodal density $\rho_c^*$, then the system obtained by flipping the sign of $\mu_{12}$ is unstable towards family demixing \emph{at the same} critical spinodal density $\rho_c^*$.\\ \indent
Lastly, let us discuss the occurence of (pure) condensation ($\theta = 0$) and demixing ($\theta = \pi/2$).  As shown in the left panel of Fig.~\ref{fig:ratio}, condensation and demixing occur, respectively, for $\zeta = 1$ and $\zeta = -c_2/c_1$. In the deterministic and uniform noise models, $\zeta$ is either $1$ or $-c_2/c_1$ when  
\begin{equation}
\mu_{11}c_1 + \mu_{12}c_2 = \mu_{12}c_1 + \mu_{22}c_2 \ .
\end{equation}
In other words, the liquid exhibits condensation and demixing when the interactions involving the two families are in balance.   In the general case of nonuniform noise amplitudes, this simple equation that identifies condensation and demixing does not apply. \\ \indent

\section{pH-induced instabilities  in the cytoplasm}\label{sec:6}

In this section, we discuss an application of the   model Eq.~(\ref{freeenergydensity}) for a complex fluid -- the  pH-driven instability of the cytoplasm of eukaryotic cells from a fluid-like to a solid-like state.    In this example, each component $i$ represents a protein.   Since the number of protein types is large (of the order $10^4$), we do not know all the entries $B_{ij}$ of the matrix of virial coefficients, and hence it is natural to consider a random model for $\mathbf{B}$.   However, since the protein-protein interactions  depend strongly on their  isoelectric points (PI),  defined as the pH at which the protein has no net charge, and since the PI  can be estimated computationally based on the amino acid sequences of the proteins  \cite{vanbogelen1999diagnosis,weiller2004modal, chan2006subcellular}, it is  desirable to group proteins into families based on their PI, and this leads to  a matrix of virial coefficients that is partially structured and random. \\ \indent
Although the cytoplasm is in general  a strongly buffered solution,  there exist cases where the cytoplasmic pH   varies significantly.    For example, it has been observed in unicellular organisms, such as, yeast cells \cite{narayanaswamy2009widespread, petrovska2014filament, munder} and bacteria \cite{setlow1980measurements}, as well as in multicellular organisms, such as shrimps \cite{busa1983intracellular}, that the  cytoplasmic pH  decreases significantly  under conditions of stress  when the cell state changes from a metabolic to a dormant state.    In yeast cells this significant decrease  of  pH  leads to the formation of macromolecular assemblies of proteins and a transition of the cytoplasm  from a fluid-like to a solid-like state \cite{narayanaswamy2009widespread, petrovska2014filament, munder}.     Another example of a phase transition under change of pH is found in the formation of skin  in mammals \cite{quiroz}.  An outward flux of cells is produced in the lower strata, which  become enucleated, flattened surface squames. Phase separation is a key component controlling this transition: biomolecular condensates are formed in the early stages of the outwards migration, which  cease to be stable as they come closer to the skin surface, and there the lowering of pH of the environment may be  the trigger of this transition \cite{quiroz}.    \\ \indent   \\ \indent
\begin{figure}[tb]
    \includegraphics[width = 7cm]{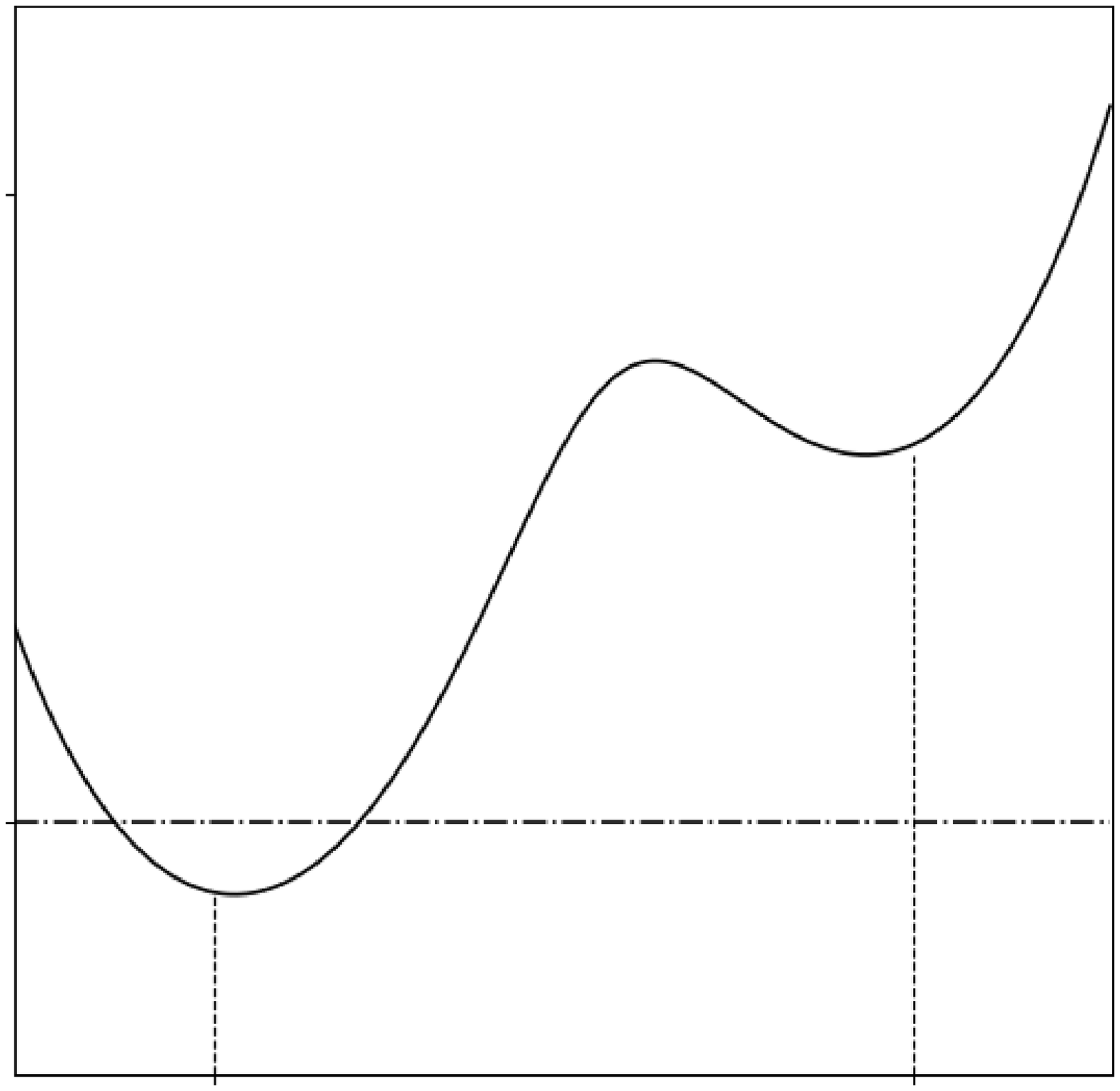}
    \quad
     \includegraphics[width = 7cm]{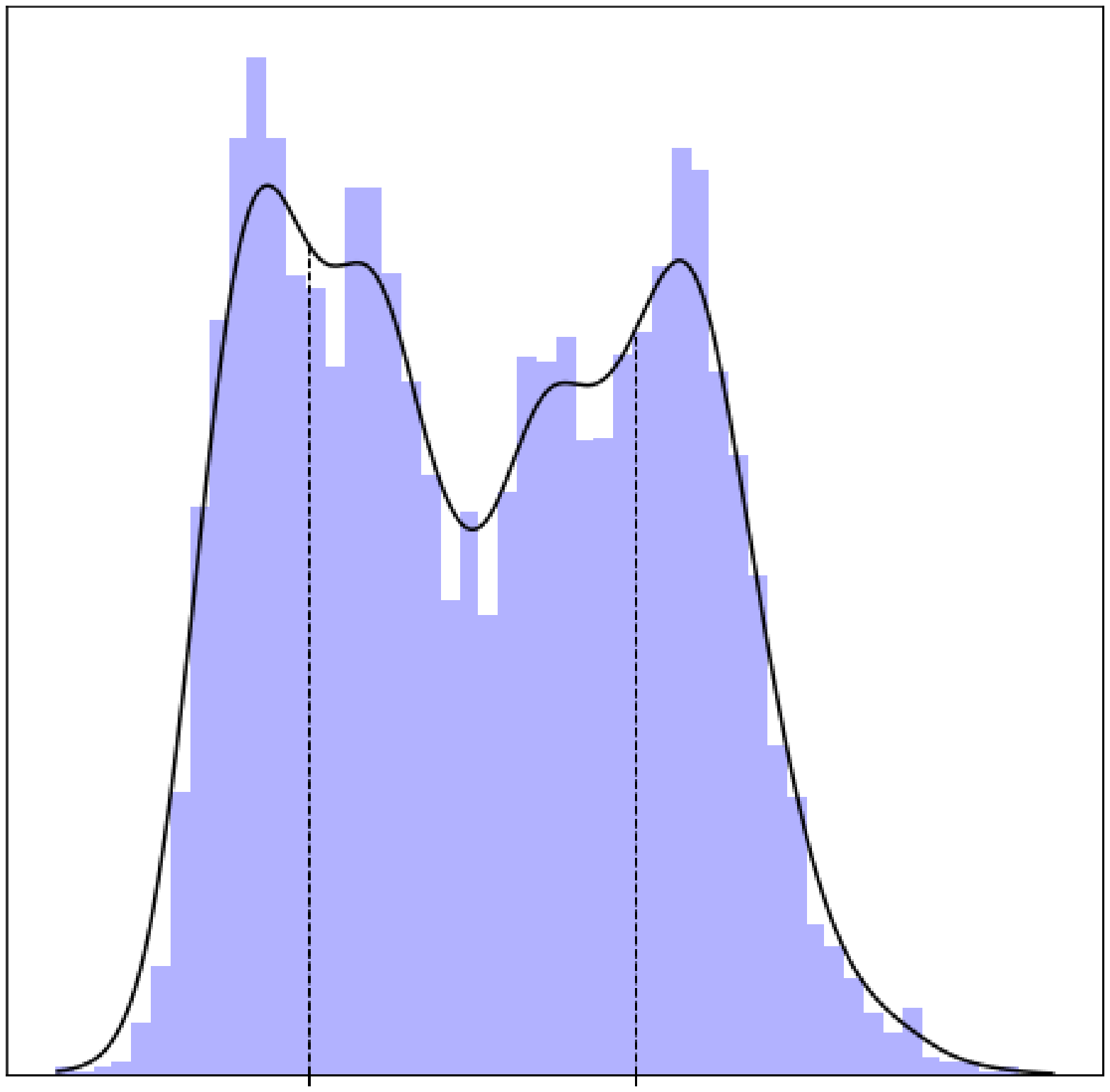}
    \put(-320,-15){pH}
     \put(-102,-15){PI}
     \put(-258,-6){$9$}
     \put(-382,-6){$5.5$}
     \put(-86,-6){$9$}
     \put(-148,-6){$5.5$}
     \put(-425,198){$\rho_c \; (10^{-3} \mathrm{nm}^{-3})$}
     \put(-426,155){$2.5$}
     \put(-426,45){$1.8$}
    \caption{Left: the critical spinodal density $\rho_c$  as a function of pH for the  model (\ref{taylorvirial}).   The dash-dot line is the value $\rho^* = 1.8\times 10^{-3} \mathrm{nm}^{-3}$ compatible with experiments of the density of proteins in the cytosol \cite{cellbiobythe}. We choose $c_1 = 0.55$ as the acidic family is more populated in living organisms. Other parameters used to produce the plot on the left are discussed in \ref{appendix:modelph}.  On the right, distribution of isoelectric points (PI) across the proteome of buddying yeast (\textit{Saccharomyces cervisiae}). Data from \cite{pidata}. }
    \label{fig:piandph}
\end{figure}
Protein interactions in solution depend strongly on their PI.    According to the   Derjaguin, Landau, Verwey, and Overbeek (DLVO) theory,  protein-protein interactions consist of an attractive part, determined by van der Waals forces, and a repulsive part, determined by electrostatic repulsion that is screened by a double layer of free ions \cite{dlvo1,dlvo2}.  When the difference between the pH and the PI of a protein increases, then the electrostatic repulsion gets larger and the average distance between proteins increases. This reduces the tendency to form aggregates. Conversely, at a pH close to their PI proteins are more likely to aggregate \cite{virialaggreg}.  \\ \indent
Since the PI strongly affects protein's interactions, we  aim to  group proteins into families based on their PI. Although the PI is a real number, we can naturally group proteins into $2$ families as the distribution of PIs across the proteome of  a large number of biological species --  ranging from unicellular organisms such as yeasts to mammals --  is  bimodal \cite{vanbogelen1999diagnosis,weiller2004modal, chan2006subcellular, pidistribcorr}.  The bimodality of the PI distribution provides the following natural way to split proteins into $2$ families: we group all proteins that have a PI smaller than   the middle point into an acidic family and all the rest into a basic family (see the right Panel of Fig.~\ref{fig:piandph}).    Without loss of generality, we choose to order the proteins so that species $1, 2, \dots, N_1 $ belong to the acidic family and proteins $N_1 + 1, \dots, N$ belong to the basic one. In this convention,  the virial matrix is a matrix with a $2 \times 2$ block structure, to which the framework of partially structured and partially random models developed  in Sec.~\ref{sec:3} applies.     \\ \indent 
Since we want to describe the instability of the cytoplasm induced by a change of pH, we  still need to determine how the entries of the virial matrix depend on the solvent's pH.
The entries of the matrix of virial coefficients $\mathbf{B}$ can be measured with,  e.g. optical experiments~\cite{experimental,scatteringvirial,crossvirial1}.  However, these  experiments have so far been done  for a handful of proteins, whereas we need to know   $B_{ii}$ and $B_{ij}$ for all proteins in the proteome as a function of pH, which to the best of our knowledge are currently not known. This motivated us to examine a simple random model for the dependence of intra-family and inter-family virial coefficients on the solvent's pH. For intra-family virial coefficients,  we consider a Taylor expansion around the average isoelectric point plus noise, i.e., 
\begin{equation}\label{taylorvirial}
    \eqalign{B_{ij} (x) = k (y_s - x)^2  + q + \xi_{ij} \;  \quad {\rm when} \quad s(i) = s(j) = s,}
\end{equation}
where $x$ is the solvent's  pH, $y_s$ is the average PI in family $s$, $k$ and $q$ are expansion coefficients, and $\xi_{ij}$ are random variables with zero mean.  For simplicity, we  use for the inter-family virial coefficients  a pH-independent constant $\mu_{12}$ plus noise $\xi_{ij}$, with $s(i) \neq s(j)$, and we set the variances of both inter- and intra-family noises $\xi_{ij}$ to be uniform and equal to $\sigma^2$. \\ \indent
In \ref{App:D1}, we estimate  the parameters $k$, $q$, $y_1$, $y_2$, and $\mu_{12}$ based on the order of  magnitude of experimentally measured protein-protein virial coefficients. Using the parameters estimated in \ref{App:D1}, the minimum eigenvalue of $\mathbf{B}$ is an outlier for all considered values of the pH, and therefore the instability of the cytoplasm is either a family demixing or condensation.     As noted in Sec.~\ref{sec:5}, the sign of $\mu_{12}$ determines whether the unstable mode is characterised by family condensation or family demixing.  Unfortunately,  we could not find data in the literature to estimate  the sign of $\mu_{12}$, and therefore we   cannot discern the exact nature of the  instability of the cytoplasm.  Nonetheless,  using Eq.~(\ref{wigner_critdens})  we can provide an estimate for the critical spinodal density $\rho_c$, which is shown in the left panel of Fig.~\ref{fig:piandph}, and since this density  is symmetrical w.r.t. the change of sign of $\mu_{12}$ it applies to both kinds of instabilities.   \\ \indent
From Fig.~\ref{fig:piandph} we observe three interesting features  for the  critical spinodal density of the cytoplasm:  i) the order of magnitude of the critical spinodal density is close to the protein density in living cells; ii) the non-monotonicity of $\rho_c$ entails the possibility of a re-entrant behavior, as highlighted in \cite{julicherph}; iii) an asymmetry in  family sizes ($c_1 \neq 0.5$) leads to an asymmetry of the critical spinodal density w.r.t. the neutral pH.  This strengthens the idea that pH can be used by living organisms as a control parameter for phase separation, and it also shows that these simple models are a promising tool to interpret biological data.  \\ \indent 
Although the model predicts that the fluid instability of the cytosol is of the family condensation or demixing type, it is interesting to investigate at which densities the  random demixing instability becomes relevant.   To investigate this, we estimate the variance $\sigma^2$ of the virial coefficients  using  fluctuations in the  volume of proteins and we use that the number of different proteins in the cytosol is  of the order $\sim 10^4$ \cite{cellbiobythe}.  From (\ref{wigner_critdens}) we obtain a critical spinodal density $\rho_c \sim \mathcal{O}(1 \; \mathrm{nm}^{-3})$ (see  \ref{app:D2}), three orders of magnitude higher than proteins concentration in the cytoplasm \cite{cellbiobythe}. This suggests that a random demixing instability due to fluctuations in excluded volume is likely not to occur in living cells. \\ \indent 

\section{Conclusions}\label{sec:conclusion}
We have determined how randomness and structure  in the  matrix of virial coefficients affect the instability of complex fluids.  In contrast with fully random models \cite{searcuesta2003}, partially structured and  random models  exhibit two types of demixing, namely, random demixing, which is akin to what has been observed in random models, and family demixing, where proteins phase separate based on the family to which they belong.   The critical spinodal density of family demixing is finite, in contrast with the critical spinodal density of random demixing that scales as $\sqrt{N}$.  Hence, family demixing is compatible with experimental findings in cell biology where demixing has been observed at finite densities.\\ \indent
We have applied the formalism of partially structured and partially random matrices of virial coefficients to the cytoplasm, which is an example of a complex fluid. Experiments in, among others,  yeast cells \cite{munder} and skin cells in mammals \cite{quiroz},  show that a change in pH can trigger a phase separation event.  In this case,  proteins can be grouped into two families according to their isoelectric point: a family of acidic proteins and a family of basic proteins.     Using a random virial matrix with a block structure, we have provided    predictions for the critical spinodal density and the nature of the instability   using only the statistical properties of inter- and intra-family protein interactions.    In particular, if the inter-family virial coefficients are positive on average, then the predicted cytoplasm instability is a family demixing, whereas if the inter-family virial coefficients are negative on average, then the instability is family condensation.  It will be interesting to compare the predicted instabilities with those observed in  experiments with, e.g., yeast cells \cite{munder}.  For example,  an intriguing question is whether the cytoplasmic instability observed in yeast experiments is a demixing or condensation transition.  \\ \indent 
We briefly discuss some of the assumptions made in the model studied in this paper.  In the definition of the   free energy model  (\ref{freeenergydensity}), we have truncated  the virial expansion of the free energy at the second order.   This raises the question whether higher order virial coefficients can be incorporated in the current formalism.    It is in principle  possible to include higher order virial expansion terms in the free energy (\ref{freeenergydensity}), see Eq.~(\ref{eq:Hessianhigherorder}) in \ref{appendix:virial}, but this would imply that off-diagonal elements of the matrix $\mathbf{H}$ depend on the densities, and this effect increases with the fluid density.  Such an amplification could stabilise the homogeneous state for high densities or could provide different mechanisms for the onset of instabilities.   Another assumption of the present model is that the  reference state is uniform, i.e. $\vec{\rho}_0 = (\rho/N, \dots, \rho/N)$.  We have chosen this reference state because it renders the calculations simpler.   A more general approach fixes the relative abundances of each species and varies the total density. This leads to the concept of dilution line, we refer the reader to \cite{Sollich2001} for a detailed discussion.   However,  the random matrix methods used in this paper are versatile enough to deal with nonhomogeneous reference states, and this will be   discussed  in a future work.   \\ \indent 
The concept of partially structured and partially random matrices of virial coefficients can be applied  in  existing models of phase separation of the cytosol.  For example,  it will be interesting to consider a complex fluid of proteins that can be charged dynamically through protonisation and deprotonisation, as done in  Ref.~\cite{julicherph} for a fluid with one component, by describing the known parts of the interaction matrix with the deterministic matrix $\mathbf{B}_{\rm det}$ and the unknown parts of the system with noise dressing the entries of $\mathbf{B}$.    Another interesting research problem is to analyse spatial positioning of droplets in a complex fluid \cite{lee2013spatial, kruger2018discontinuous}.
Taken together,  the theory of partially structured and partially random matrices  provides a versatile tool for understanding the behaviour of complex fluids.\\

\ack
GC would like to thank Francesco Coghi and Gianmichele Di Matteo for insightful discussions. IN thanks Omar Adame-Arana, Pablo Sartori and Francesco Turci for fruitful discussions. GC is supported by the EPSRC Centre for Doctoral Training in Cross-Disciplinary Approaches to Non-Equilibrium Systems (CANES, EP/L015854/1).

\appendix

\section{Virial expansion}\label{appendix:virial}

In this appendix we review a classical argument -- see \cite{Zamponinotes, landau5, mmtheory} -- to derive the virial expansion of the free energy of an interacting mixture of $N$ chemical species, and comment on how including higher order terms in the expansion influences the Hessian of the free energy.  \\ \indent
We consider a system composed of $M$ different particles of $N$ distinct species. We use indexes $a,b \in\{1, \ldots, M\}$ for particles and indexes $i,j \in \{1, \ldots, N\}$ for species. We denote by $M_i$ the number of particles of species $i$ present in the mixture, satisfying $\sum_i M_i = M$. We start from the Hamiltonian:

\begin{equation}
    \mathcal{H}(\bm{q}, \bm{p}) = \sum\limits_a^M \frac{p_a^2}{2m} + \sum_{a, b} v_{a b}(|q_a - q_b|) \ ,
\end{equation}
where the first term is the kinetic energy, the second term the interaction energy given by a sum of pairwise potentials and $(\bm{q}, \bm{p})$ are canonical coordinates and momenta, respectively. To account for the different nature of species composing our mixture, we let the pairwise potentials $v_{a b}(|q_a - q_b|) = v_{ij}(|q_a - q_b|)$ depend only on the species $i$ and $j$ of particles $a$ and $b$ respectively. The canonical partition function of such a system reads 

\begin{equation}
    Z = \frac{1}{h^{3M}\prod_i M_i!} \int \! \mathrm{d}\bm{q}\mathrm{d}\bm{p} \; \mathrm{e}^{-\beta \mathcal{H}(\bm{q},\bm{p})} \ ,
\end{equation}
where $h$ is the Planck constant. The partition function factorises in the kinetic part and the interacting part: the former can be evaluated via gaussian integration, while the latter is non-trivial and in general cannot be computed. We call the configurational integral appearing in the partition function $Z_M$:

\begin{equation}
    Z_M = \frac{1}{h^{3M}\prod_i M_i!}\int \! \mathrm{d}\bm{q} \; \mathrm{e}^{-\beta \sum_{a,b} v_{ij}(|q_a - q_b|)} \ .
\end{equation}
If the mixture was \emph{ideal}, which means non-interacting, the configurational integral would simply be $Z_M^{id} = V^M$, where $V$ is the volume. In order to perturbatively study small deviations from ideality we multiply and divide by $Z_M$, so that we can express the free energy $F = - \beta^{-1} \log Z$ as

\begin{equation}
    F = F_{kin} - \frac{1}{\beta}\log\left(\frac{V^M}{h^{3M}\prod_i M_i!}\frac{ Z_M}{V^M} \right) = F_{id} - \frac{1}{\beta}\log\left(\frac{ Z_M}{V^M} \right) = F_{id} + F_{int} \ ,
\end{equation}
where $F_{kin} = \sum_{i} 3 M_{i} T\log\Lambda$ is the kinetic free energy, $\Lambda = \sqrt{2\pi\beta \hbar^2/m}$ is the De Broglie thermal wavelength and $F_{id} = F_{kin} + \sum_{i}M_{i} T (\log M_i/V - 1)$ is the free energy of a non interacting mixture. \\ \indent 
Calling $r_{ab} = |q_a - q_b|$, we note that the Gibbs weight of a pairwise potential $\mathrm{e}^{-\beta v(r_{ab})}$ goes to $1$ for $r \to \infty$, because the potential falls off to $0$. In the spirit of an expansion involving a small quantity, the aforementioned consideration motivates the introduction of the Mayer function $\mathrm{e}^{-\beta v(r_{ab})}-1 $, which goes to $0$ for large $r$. We have:

\begin{equation}\label{configurational}
    F_{int} = - \frac{1}{\beta}\log\left( 1 + \frac{1}{V^N}\int \! \mathrm{d}q^{3M} \; (\mathrm{e}^{-\beta \sum_{a,b} v_{ij}(|q_a - q_b|)} - 1)\right)\ .
\end{equation}
We expect that deviations from ideality are small when the number densities $M_i/V$ are small enough. In this regime, the dominant configurations contributing to (\ref{configurational}) are the ones in which only $k$ particles are interacting. This is because the probability to choose $k$ particles is proportional to $\rho_{i_1} \rho_{i_2} \dots \rho_{i_k}$, where $i_1, \dots, i_k$ are the species of particles $1, \dots, k$ respectively. One can thus arrange the sum over these configurations as a power series in the densities, called the cluster expansion:

\begin{equation} \label{eq:clusterexpansion}
    F_{int} \approx \frac{V}{\beta} \sum_{k = 2}^{k_{max}} \sum\limits_{i_1, \dots, i_k} B_{i_1 \dots i_k}(T) \rho_{i_1} \dots \rho_{i_k}  \ ,
\end{equation}
where $B_{i_1 \dots i_k}$ are called virial coefficients and we included terms up to order $k_{max}$. We can express the virial coefficients from (\ref{configurational}) using the approximation $\log(1 + x) \approx x$. For second order virial coefficients this leads to
    
\begin{equation} 
     B_{ij}(T) =  -\frac{1}{2} \int \! \mathrm{d}\bm{r} \; (\mathrm{e}^{-\beta v_{ij}(\mathbf{r})} - 1) \ .
\end{equation}
Truncating the virial expansion to second order, we arrive at the following expression for the free energy density

\begin{equation}\label{eq:fullfreeenergy}
    \beta \tilde{f} = \frac{\beta F}{V} =  3 \log \Lambda \sum_i  \rho_i  + \sum_i \rho_i(\log \rho_i -1) + \frac{1}{2} \sum_{ij} \rho_i B_{ij} \rho_j \ .
\end{equation}
We note that expression (\ref{eq:fullfreeenergy}) differs from (\ref{freeenergydensity}) only because of the kinetic free energy density $3 \log \Lambda \sum_i  \rho_i $. It is common practice to neglect this term, as it does not affect the phase behaviour \cite{SollichCates}. In fact, instabilities are tied to properties of the Hessian, in which second derivatives kill linear terms of the free energy. This justifies the use of (\ref{freeenergydensity}) to study instabilities of complex fluids.

\subsection{The effect of higher order non-idealities on the Hessian of the free energy} 

In this subsection we consider a free energy density of the form 

\begin{equation}
     f  = \frac{1}{V}(F_{id} + F_{int}) \ , 
\end{equation}
where $F_{id}$ is the free energy of an ideal mixture of $N$ components and $F_{int}$ is of the form (\ref{eq:clusterexpansion}). We have set $\beta =1$ for simplicity. The Hessian of such a free energy can be expressed as

\begin{equation}\label{eq:Hessianhigherorder}
    \beta H_{ij} = \frac{\partial^2 F}{\partial \rho_i \partial \rho_j} = \delta_{ij} \frac{1}{\rho_i} + \sum_{k = 2}^{k_{max}} k(k-1)\sum\limits_{\ell_1, \dots, \ell_{k-2}} B_{\ell_1 \dots \ell_{k-2} ij} \  \rho_{\ell_1} \dots \rho_{\ell_{k-2}}\ . 
\end{equation}
We note one important difference between the Hessian in  (\ref{eq:Hessianhigherorder}) and its counterpart (\ref{eq:Hessian}) obtained from a truncation to the second order of the virial expansion: diagonal terms in the former depend on the densities $\vec{\rho}$ of the state we are considering. We can understand the effect of this feature on the spectrum of $\mathbf{H}$ by considering a dilute uniform reference state $\vec{\rho_0} = (\rho/N , \dots, \rho/N)$ and increasing the total density $\rho$. When we include only second order terms in $F_{int}$, an increase in $\rho$ makes the spectrum of $\mathbf{H}$ translate rigidly; on the other hand, when we include more terms in $F_{int}$ an increase in $\rho$ not only translates the spectrum of $\mathbf{H}$ towards the negative semiaxis, but it amplifies the effect of higher order terms. Such an amplification could have the effect of stabilising homogeneous phases with high total density $\rho$, as expected from realistic fluid models, or could provide different mechanisms for the onset of instabilities. \\ \indent

\section{Spectrum of block matrices}\label{appendix:spectrumblock}
Let $ \mathbf{B}$ be a symmetric $N \times N$ random matrix with a $F\times F$ block structure in the following sense:   the entries $B_{ij}$ are independent random variables with  finite means $\langle B_{ij} \rangle = \mu_{s(i)s(j)}$ and variances $\mathrm{Var}(B_{ij}) = \sigma^2_{s(i)s(j)}$, where the function $s(i)$ returns the family $s \in \{1, \dots, F\}$ to which the $i$-th index belongs. \\ \indent
In this appendix, we derive a set of equations that determine the empirical spectral density 
\begin{equation}
    \nu(\lambda) = \lim \limits_{N \to \infty} \frac{1}{N}\sum^N_{i=1} \delta(\lambda_i - \lambda)
\end{equation}
of the eigenvalues $\lambda_i$ of $\mathbf{B}$, the eigenvalue outliers $\lambda_{\rm isol}$, and the distribution of the entries  of the eigenvectors associated  with either the outlier eigenvalues or the eigenvalues at the edge of the spectral density. \\ \indent
It is convenient to consider a rescaled version of $\mathbf{B}$, which we will call $\tilde{\mathbf{B}}$, whose entries have means $\langle \tilde{B}_{ij} \rangle = \mu_{s(i)s(j)}/N$ and variances $\mathrm{Var}(\tilde{B}_{ij}) = \sigma_{s(i)s(j)}^2/N$. This choice is such that both outliers and the support of the spectral density $\nu(\lambda)$ of $\tilde{\mathbf{B}}$ are of order $\mathcal{O}(1)$ for large $N$, and they are connected to those of $\mathbf{B}$ as follows:
\begin{eqnarray}
    \nu_{B}(\lambda) = \frac{1}{\sqrt{N}}\nu_{\tilde{B}}(\lambda/\sqrt{N}) \\
    \lambda_{\rm isol}(B) \; (\mathbf{D},\mathbf{C})=  \lambda_{\rm isol}(\tilde B)\; (N\mathbf{D},\sqrt{N}\mathbf{C}) \, ,
\end{eqnarray}  
where  $\mathbf{D}$ and $\mathbf{C}$ are $F\times F$ symmetric matrices containing the means $\mu_{st}$   and the standard deviations $\sigma_{st}$ defined in section \ref{sec:3}.

\subsection{A reminder of some  equalities for block matrices}\label{schur}
In this section, we review a few equalities for the inverse and the determinant of a block matrix.   Let $\mathbf{M}$ be an invertible $N \times N$ block matrix,
\begin{equation}
    \mathbf{M} =  \left(\begin{array}{@{}c|c@{}}
    \mathbf{a} & \mathbf{b}   \\\hline
    \mathbf{c} & \mathbf{d} 
  \end{array}\right) \; .
\end{equation}
then the Schur formula for the inverse of $\mathbf{M}$ states \cite{tao2012topics} 
\begin{equation} \label{eq:schur}
 \mathbf{M}^{-1} = 
  \left(\begin{array}{@{}c|c@{}}
    \mathbf{s_d^{-1}} & \mathbf{-s_d^{-1}bd^{-1}}   \\\hline
    \mathbf{-d^{-1}cs_d^{-1}} & \mathbf{s_a^{-1}} 
  \end{array}\right) \; ,
\end{equation}
where 
\begin{equation}
     \mathbf{s_a} =  \mathbf{d} -  \mathbf{c} \mathbf{a}^{-1} \mathbf{b} \; ,
     \end{equation}
and 
\begin{equation}
     \mathbf{s_d} =  \mathbf{a} -  \mathbf{b} \mathbf{d}^{-1} \mathbf{c} \; .
\end{equation}

We will also use the  formula
\begin{equation}\label{eq:determinantblock}
    \mathrm{det}(\mathbf{M}) = \mathrm{det}(\mathbf{a} - \mathbf{b} \mathbf{d}^{-1} \mathbf{c} )\, \mathrm{det}(\mathbf{d})
\end{equation}
for the determinant of a block matrix.  

\subsection{The resolvent}\label{app:theresolvent}

A standard approach in random matrix theory to determine the spectrum of a large random matrix is based on the resolvent $G(z)$, defined as
\begin{equation}
    G(z) = \frac{1}{z \mathbb{1} - \mathbf{B}} \; ,
\end{equation}
for all values $z \in \mathbb{C} / \left\{\lambda_1,\lambda_2,\cdots,\lambda_N\right\}$, and where $\mathbb{1}$  is the identity matrix of size $N$.   Let 
\begin{equation}\label{nromtraceres}
  \mathcal{G}(z) = \lim_{N\rightarrow \infty}  \frac{1}{N} \mathrm{Tr} G(z) 
  \end{equation} 
  be the trace resolvent, 
 then the spectral density follows readily  from~\cite{vivobook}
\begin{equation}\label{eq:specDens}
    \nu(\lambda) = \lim_{\epsilon \to 0} \frac{1}{\pi} \mathrm{Im} \; \mathcal{G}(\lambda - \mathrm{i} \epsilon) \; .
\end{equation} 

The trace resolvent   $\mathcal{G}(z)$ of a random matrix  can be determined from the Schur formula (\ref{eq:schur}), see \cite{bouchaud}.   In what follows we  use this method to determine  the trace resolvent of the block matrix $\mathbf{B}$.  Let us represent  $z\mathbb{1} - \tilde{\mathbf{B}} $ as  a $2 \times 2$ block  matrix
\begin{equation}
    z \mathbb{1} - \tilde{\mathbf{B}} = 
  \left(\begin{array}{@{}c|c@{}}
    z - \tilde{B}_{11} & \vec{B}_1  \\\hline
    \vec{B}_1^T & (z \mathbb{1}-\tilde{\mathbf{B}})^{(1)} 
  \end{array}\right) \; ,
\end{equation}
where $\vec{B}_1$ is the $(N-1)$-dimensional vector of components $\tilde{B}_{1j}$ with $j \neq 1$. We use $\mathbf{M}^{(i)} $ to denote the $(N-1) \times (N-1)$ principal submatrix obtained from $\mathbf{M}$ by eliminating the $i$-th row and  $i$-th column. Applying the Schur formula to $(z\mathbb{1}-\tilde{\mathbf{B}})^{-1}$, we can express the diagonal  $(1,1)$-element as
\begin{equation}
    \frac{1}{G_{11}} = z - \tilde{B}_{11} - \sum^N_{j=1}\sum^N_{k = 2} \tilde{B}_{1j} G^{(1)}_{jk} \tilde{B}_{k1} \; , \label{eq:G11}
\end{equation} 
where we used $G^{(1)} = 1/(z\mathbb{1} - \tilde{\mathbf{B}}^{(1)})$.
Permuting the rows and columns of the matrix $\tilde{\mathbf{B}}$ and applying the Schur formula gives an expression similar to (\ref{eq:G11}) where index $1$ is replaced by index $i$, viz., 
\begin{equation}
    \frac{1}{G_{ii}} = z - \tilde{B}_{ii} - \sum^N_{j=1}\sum^N_{k =1 (k\neq i)} \tilde{B}_{ij} G^{(i)}_{jk} \tilde{B}_{ki} \; . \label{eq:Gii}
\end{equation} 
For large values of  $N$, we can neglect the $\tilde{B}_{ii}$ term in (\ref{eq:Gii}) as it scales as $\mathcal{O}(1/N)$  with the system size.  In addition, in the \ref{appendix:offdiag} we show that the  scaling of the off-diagonal elements of the resolvent  with respect to $N$ is  subleading with respect to the scaling of the diagonal elements, and therefore  
\begin{equation}\label{eq:largenG11}
\frac{1}{G_{ii}} = z - \sum^N_{j=1 (j\neq i)}\tilde{B}^2_{ij} G_{jj}^{(i)} \; .
\end{equation} 
Because of the law of large numbers it holds that for large values of $N$ the sum in the right-hand side of (\ref{eq:largenG11}) is equal to its average value and therefore $G_{ii}$ is a deterministic variable in the limit of $N$ large.  In particular, we find that $G_{ii}$ only depends on the family $s(i)$ to which the index $i$ belongs, allowing us to make the simplification $G_{ii} = G_{s(i)}$.   In addition to that, using that  $G_{jj}^{(i)}= G_{jj}(1+\mathcal{O}(1/N)) = G_{s(j)} (1+\mathcal{O}(1/N))$,  we arrive  at 
\begin{equation} \label{setdiagres}
    \frac{1}{G_{s}} = z - \sum^F_{t=1} c_t \sigma_{st}^2G_{t} \;,
\end{equation} 
with $c_t = \lim_{N \to \infty} N_t/N$, which is also (\ref{systemresolvent}) in the main text. Solving the set of $F$ equations (\ref{setdiagres}) towards the $F$ variables $G_s$, one obtains for large $N$ the trace resolvent
\begin{equation} 
\mathcal{G} =\sum^F_{t=1}  c_t  G_t \; , \label{eq:A16}
\end{equation}
and thus also the spectral density through (\ref{eq:specDens}).

\subsection{Off-diagonal elements of the resolvent}\label{appendix:offdiag}
In this section we show that the  off-diagonal elements $G_{ij}$ with $i\neq j$ scale as $1/N$.    In particular, we show that both  the first and second moments  of  $G_{ij}$ scale as $1/N$. \\ \indent
We start from the adjugate representation of the inverse  of a matrix to express $G_{ij}$ as
\begin{equation}
    G_{ij} = \frac{\det(z\mathbb{1}-\tilde{\mathbf{B}}^{(i,j)})}{\det(z\mathbb{1}-\tilde{\mathbf{B}})} \; ,
\end{equation}
where $\tilde{\mathbf{B}}^{(i,j)}$ is obtained from $\tilde{\mathbf{B}}$ by removing the $i$-th row and the $j$-th column.  Using the formula  (\ref{eq:determinantblock}) for the determinant of a block matrix  we obtain
\begin{eqnarray}\label{offdiag}
  \fl   G_{ij} 
  =  \frac{ - \tilde{B}_{ij} - \sum\limits_{k,l (\neq i,j)}^N  \tilde{B}_{ik} G^{(i)(j)}_{kl} \tilde{B}_{jl}  }{  \left[z- \tilde{B}_{ii}-\sum\limits_{k,l (\neq i,j)}^N  \tilde{B}_{ik} G^{(i)(j)} _{kl} \tilde{B}_{il}\right] \left[z-  \tilde{B}_{jj} -\sum\limits_{k,l(\neq i,j)}^N  \tilde{B}_{jk} G^{(i)(j)}_{kl} \tilde{B}_{jl}\right] - \left[\sum\limits_{k,l(\neq i,j)}^N  \tilde{B}_{ik} G^{(i)(j)}_{kl} \tilde{B}_{jl}\right]^2} \, , \nonumber\\
\end{eqnarray}
where $G^{(i)(j)}$ denotes the resolvent of the matrix $\mathbf{B}^{(i)(j)}$ obtained form $\mathbf{B}$ by removing the $i$-th row and column and the $j$-th row and column.  For large $N$, we have that  $G^{(i)(j)}_{kl} = G_{kl}(1+\mathcal{O}(1/N))$.  \\ \indent 
Suppose now that the average of $G_{kl}$ scales as $N^{-\delta_1}$ and its variance scales as $N^{-\delta_2}$, where $\delta_1,\delta_2\in\mathbb{R}$. Equation (\ref{offdiag}) provides us with a self-consistent equation for the exponents $\delta_1$ and $\delta_2$.  Solving this equation we find that  $\delta_1=\delta_2=1$. This implies that the average value and the variance of $G_{ij}$ are both of the order  $\mathcal{O}(1/N)$.   \\ \indent
We remark that analogous arguments hold also for the  resolvent $G^{(i)}$ of the matrix $\mathbf{B}^{(i)}$, in particular $G^{(i)}_{jk}$ is negligible for large $N$ w.r.t the diagonal elements $G_{jj}^{(i)}$.

\subsection{Outlier eigenvalues and the components of eigenvectors}\label{appoff}
We derive a set of equations solved by the entries $v_i$ of  eigenvectors associated with either eigenvalue outliers or eigenvalues located at the boundary of the continuous part of the spectrum of block matrices.  The  general approach we follow is based on \cite{izaak} that deals with sparse random matrices, but as we will see, dealing with dense matrices has some advantages and lead  to some analytical simplifications.   \\ \indent
First we  establish a connection between  the eigenvector components and the off-diagonal elements of the resolvent $G(z)$. Let $\lambda$ be an eigenvalue of $\tilde{\mathbf B}$ with an algebraic  multiplicity equal to $1$, then the resolvent $G(z)$ has a simple pole at $\lambda$ and \cite{izaak}
\begin{equation}\label{limit}
    \lim_{\eta \to 0 } \eta G(\lambda - \eta) = \vec{v} \; \vec{v}^{T} \; , 
\end{equation}
where $\eta$ is a small complex number and $\vec{v}$ is the normalised eigenvector associated with $\lambda$.    It follows from (\ref{limit}) that  the $i$-th component of $\vec{v}$ is given by
\begin{equation}\label{linkreseigen}
    v_i = \lim_{\eta \to 0} \eta \frac{\sum_{j=1}^N G_{ij}(\lambda - \eta)}{\vec{v} \cdot \vec{1}} \; ,
\end{equation}
where we have used $\vec{1}$ for the vector with all components equal to $1$.  Equation  (\ref{linkreseigen}) links the components of eigenvectors  to the off-diagonal entries of the resolvent.    \\ \indent
Note that the parameter $\eta$ in (\ref{limit})-(\ref{linkreseigen}) has to be much smaller than the separation between eigenvalues.   Therefore, in the limit of $N\gg 1$, (\ref{limit})-(\ref{linkreseigen}) are only useful for eigenvalue outliers or eigenvalues at the edge of the continuous spectrum.\\ \indent
The off-diagonal entries of the resolvent solve a set of self-consistent equations that we derive now.   Applying  the Schur formula to the resolvent $G(z)$, we obtain 
\begin{equation}
    G_{ij} = G_{ii} \sum^N_{k=1  (k \neq i)} \tilde B_{ik} G_{kj}^{(i)} \; ,
\end{equation}
where  we  mean again  by  $G^{(i)}$  the principal submatrix of  $G$ obtained  by removing the $i$-th row and the $i$-th column. Summing over index $j$ we obtain
\begin{equation}
    \sum_{j = 1}^N G_{ij} = G_{ii}\left( 1 + \sum^N_{j=1} \sum_{k=1 (k \neq i)}^N  \tilde B_{ik} G_{kj}^{(i)} \right) \; , \label{eq:SumOff}
\end{equation}
and consequently substituting   (\ref{eq:SumOff}) in   (\ref{linkreseigen})  we get
\begin{equation}
    v_i = \lim_{\eta \to 0} \eta \frac{ G_{ii}(\lambda - \eta)}{\vec{v} \cdot \vec{1}} + G_{ii}\sum^N_{k=1 (k \neq i)}  \tilde B_{ik} \lim_{\eta \to 0} \eta \frac{  \sum^N_{j =1 (j\neq i)} G_{kj}^{(i)}(\lambda - \eta)}{\vec{v} \cdot \vec{1}} \; . 
\end{equation}
The first term converges to zero for large $N$ and 
\begin{equation}\label{eq:cavityvectorcomponent}
v_k^{(i)} =  \lim_{\eta \to 0} \eta \frac{\sum^N_{j=1 (j \neq i)} G_{kj}^{(i)}(\lambda - \eta)}{\vec{v}^{(i)} \cdot \vec{1}} \approx \lim_{\eta \to 0} \eta \frac{\sum^N_{j=1 (j \neq i)} G_{kj}^{(i)}(\lambda - \eta)}{\vec{v} \cdot \vec{1}} \; 
\end{equation} 
is identified as the $k$-th element of the eigenvector associated with  $\lambda$ of the submatrix $\tilde{\mathbf{B}}^{(i)}$. This identification is consistent only if $\lambda$ is an eigenvalue of both $\tilde{\mathbf{B}}$ and $\tilde{\mathbf{B}}^{(i)}$, which applies for eigenvalue outliers and the edge of the spectral density when $N$ is large. In the last passage of (\ref{eq:cavityvectorcomponent}) we have  used the law of large numbers to  identify $\vec{v}^{(i)} \cdot \vec{1}$ with $\vec{v} \cdot \vec{1}$.   From (\ref{eq:cavityvectorcomponent}) it follows that  we can  express $v_i$ in terms of $v_k^{(i)}$, viz.,
\begin{equation}
v_i =   \lim_{\eta \to 0 }G_{ii}(\lambda - \eta) \sum\limits^N_{k =1 (k\neq i)} \tilde B_{ik} v_k^{(i)} \ .
\end{equation}
The difference between $v_k$ and $v_k^{(i)}$ decreases when $N$ increases, as can be seen from comparing (\ref{linkreseigen}) with  (\ref{eq:cavityvectorcomponent}) and noting that the law of large numbers holds for both  the numerators and the denominators. With the relabelling $k \to j$, we arrive at 
\begin{equation}\label{eigvnoavg}
    v_i = \lim_{\eta \to 0 }G_{ii}(\lambda - \eta) \sum\limits^N_{j =1 (j\neq i)} \tilde B_{ij} v_j \; .
\end{equation}
We remark that (\ref{eigvnoavg}) applies to the entries $v_i$ of eigenvectors associated with eigenvalue outliers $\lambda=\lambda_{\rm isol}$ and  eigenvectors associated with eigenvalues $\lambda$ located at the edge of the continuous spectrum. \\ \indent
If $\lambda=\lambda_{\rm isol}$, then the expected values $\langle v_j\rangle \neq 0$.   Therefore, we can apply the law of large numbers to (\ref{eigvnoavg}) to get $v_i = V_{s(i)}$, where the $V_s$ are deterministic variables that solve (\ref{systemeigenvectors}).   One can verify that (\ref{systemeigenvectors}) only admits a nontrivial solution, i.e.~$V_s\neq 0$, when $\lambda=\lambda_{\rm isol}$.  Hence, the locations of the eigenvalue outliers can be obtained by setting the determinant of the linear system (\ref{systemeigenvectors}) equal to zero.   \\ \indent
If $\lambda$  is set equal to one of the edges of the continuous spectrum, then the means of $v_i$ are equal to zero.   In this case the central limit theorem applies and  consequently the $v_i$ are Gaussian random variables with zero mean and variances $\Delta_{s(i)}$ that only depend on the family to which the index $i$ belongs.    From (\ref{eigvnoavg}) it follows that the variances solve (\ref{eq:systemvarianceseigvec}).  Also, the edge of the spectrum $\lambda_{\rm b}$ can be obtained by finding the values of $\lambda$ for which (\ref{eq:systemvarianceseigvec}) admits a nontrivial solution.   For this, one can again find the values of $\lambda$ for which the  determinant of the linear system, in this case given by (\ref{eq:systemvarianceseigvec}), is equal to zero.

\section{Unstable mode due to outliers for $F=2$}\label{appendix:f2spinodal}

In this appendix we characterise explicitly the unstable mode of a complex fluid composed of $2$ families of interacting species.  We consider three cases depending on the assumptions made on the interaction matrix $\mathbf{B}$:  i) deterministic interactions; ii) random interactions with uniform variances; iii) random interactions with family-dependent variances.   Note that these are also the three cases illustrated in Fig.~\ref{fig:spectra}.  \\ \indent We focus on fluid instabilities that are governed by  an isolated eigenvalue $\lambda_{\rm isol}$.   Since eigenvectors are defined up to a proportionality constant, we can, for values $V_2\neq 0$, cast  the unstable mode  into the form $v^- = (\zeta, \dots, \zeta, 1, \dots, 1)$, where $\zeta = V_1/V_2$.   In what follows we determine  $\zeta$ as a function of the system parameters.  

\subsection{Deterministic case}
When interactions are deterministic, a complex fluid composed of $F=2$ families is unstable when the lowest of the two non-zero eigenvalues of $\mathbf{B}$ are negative.  We get for $\zeta$ the expression
\begin{equation}\label{eq:zetadeterministic}
    \zeta =  \frac{ \mu_{11}c_1 - \mu_{22}c_2}{2\mu_{12}c_1} - \frac{\sqrt{(\mu_{11}c_1 - \mu_{22}c_2)^2 + 4c_1 c_2  \mu_{12}^2}}{2\mu_{12}c_1} \; .
\end{equation}
Remarkably, the sign of $\zeta$ depends only on the sign of $\mu_{12}$, i.e.,  $\mathrm{sgn}(\zeta) = - \mathrm{sgn}(\mu_{12})$.

\subsection{Uniform noise}
When interactions are noisy, the instability is either due to the lower edge of the continuous part of the spectrum or due to  one of the outliers. When noises are uniform, the latter case is particularly simple. In fact, the solubility condition of (\ref{systemeigenvectors}) is
\begin{equation}\label{outwigner}
    \mathcal{G}(\lambda_{\rm isol}) = \frac{1}{\gamma_{\rm isol}} \; .
\end{equation}
Substituting   (\ref{outwigner})  back into (\ref{systemeigenvectors}) we obtain that $V_1$ and $V_2$ solve the same equations as those solved by  the corresponding eigenvalue in the deterministic case and therefore   $\zeta$ is also given by (\ref{eq:zetadeterministic}). \\ \indent

\subsection{Family-dependent noise}\label{app:c3}
Contrary to the uniform case, when the noise in the virial matrix is family dependent and the instability is due to an outlier $\lambda_{\rm isol}$, the unstable mode depends also on the variances of the noise. When $F=2$, we can express $\zeta$ from (\ref{systemeigenvectors}) as
\begin{equation}\label{zetag}
    \zeta = \frac{G_1(\lambda_{\rm isol}) c_2 \mu_{12}}{1 - G_1(\lambda_{\rm isol}) c_1 \mu_{11}} \; . 
\end{equation}
From (\ref{setdiagres}), one can prove that $G_1(z)$ solves a quartic equation

\begin{equation}\label{quartic}
\fl \eqalign{\sigma_{11}^2 c_1^2 (\sigma_{12}^4 - \sigma_{11}^2\sigma_{22}^2)G_1^4(z) + c_1 (2\sigma_{11}^2\sigma_{22}^2- \sigma_{11}^2 \sigma_{12}^2 -\sigma_{12}^4) z G_1^3(z)+ \cr + [z^2(\sigma_{12}^2-\sigma_{22}^2) + (c_1 - c_2)\sigma_{12}^4- 2c_1\sigma_{11}^2\sigma_{22}^2]G_1^2(z) + (2 \sigma_{22}^2 -  \sigma_{12}^2)zG_1(z) -  \sigma_{22}^2 =0 \; . }
\end{equation}
Numerical calculations show that if $\lambda_{\rm isol}$ is the lowest outlier of $\tilde{\mathbf{B}}$, the sign of $\zeta$ depends only on the sign of $\mu_{12}$ as in previous cases: $\mathrm{sign}(\zeta) = - \mathrm{sgn}(\mu_{12})$.

\section{Parameters used  for the  model of pH-induced phase transitions in the cytosol used  in Fig.~\ref{fig:piandph}}\label{appendix:modelph}  
We first discuss the parameters used in Fig.~\ref{fig:piandph}  for the model  for pH induced phase transitions in the cytosol described in Sec.~\ref{sec:6}, and second, we estimate the critical spinodal density at which volume fluctuations destabilise  a fluid of hard spheres.

 \subsection{Estimated parameters for the cytosol}\label{App:D1}
Reference~\cite{pidata} contains the isoelectric points of the proteomes of several organisms, among which humans and buddying yeast.   For example, in Fig.~\ref{fig:piandph} we have plotted the distribution of isoelectric points in the latter.   Since the distribution is bimodal, we can  use the  two peak values  of the distribution to estimate the average PI in each family.   For budding yeast this gives $y_1 = 5.5$ and $y_2 = 9$.   Furthermore, from Fig.~\ref{fig:piandph} we estimate the fraction of species in the acidic family to be $c_1 = 0.55$.    
\\ \indent
In \cite{experimental,scatteringvirial} the virial coefficients of some proteins have been measured experimentally.   From their results, we identify a plausible physical range of the average intra-family virial coefficients between $ - 10^3\; {\rm nm}^3$ and + $10^3 \; {\rm nm}^3$.   To reproduce this range of values, in the model given by (\ref{taylorvirial})  we set the parameter values $k = 10^2 \; {\rm nm}^3$ and $q = - 10^3 \; {\rm nm}^3$.  Experimental results of \cite{crossvirial1} indicate that on average the cross virial coefficients are one order of magnitude less than the intra-species virial coefficients, and therefore we set $\mu_{12} = \pm 250 \; {\rm nm}^3$ -- we remark that the sign of $\mu_{12}$ has no effect on the critical spinodal density, but it would be crucial to distinguish between family demixing and family condensation. We therefore have to give up the description of the unstable mode for lack of data. \\ \indent
By assuming that proteins are hard spheres, the variance $\sigma^2$ of the  noise variables  can be estimated  from  the fluctuations in  the  protein volumes.    Indeed, for hard spheres an explicit expression for $B_{ij}$ is known.   If species $i$ and $j$ have radii $r_i$ and $r_j$, respectively, then  \cite{virialpolydispersehardspheres}:
\begin{equation}
    B_{ij} = \frac{2}{3} \pi (r_i + r_j)^3 \; . 
\end{equation}
In particular, when $i=j$, then $B_{ii} = 4 V$, where $V$ is the volume of a sphere of radius $r_i$.  Hence, fluctuations in the volumes of hard spheres lead to fluctuations in the virial coefficients $B_{ij}$. \\ \indent
We can estimate the standard deviation $\sigma$ of $B_{ij}$ from the variance $\sigma_V$ of the protein volume $V$ with the formula 
\begin{equation}
\sigma = 4 \: \sigma_V \; . \label{eq:sigmaV}
\end{equation}

To estimate the variance $\sigma^2_V$ of the volume fluctuations of proteins, we start from the distribution of the number of amminoacids in proteins.  The standard deviation of this distribution is of the order  $\sim 10^2$ \cite{cellbiobythe}.  Since the weighted average of amminoacids' mass is $\sim 110 \; {\rm Da}$, where $1 \; {\rm Da} = 1 \; {\rm g} / 1 \; {\rm mol} $, the standard deviation in protein mass is of the order $10^4$ Da.  For folded proteins, we have that  \cite{masstovolume}
\begin{equation}
    V ({\rm nm}^3) = \frac{1}{\phi_{\rm p}} \times M({\rm Da}) \; ,
\end{equation}
where 
\begin{equation}
    \phi_{\rm p} = 0.825 \left(\frac{{\rm Da} }{{\rm nm}^3}\right)
\end{equation}
is the average density of a protein.   This phenomenological relation gives a standard deviation for the volume of the order $\sigma_V\sim 10-10^2 \; {\rm nm}^3$ and thus from (\ref{eq:sigmaV}) $\sigma \sim 10-10^3 \: {\rm nm}^3$.   To generate the plot in Fig.~\ref{fig:piandph}, we have used $\sigma = 100 \: {\rm nm}^3$.

\subsection{Critical  spinodal density $\rho_c$ for a complex fluid of hard spheres}\label{app:D2}

We can estimate the number of protein species in a cell to be approximately the number of protein coding genes, which is $N \sim 10^4$ \cite{cellbiobythe}. Recalling that for random demixing $\rho_c = \sqrt{N}/\sigma$, we get the estimate $\rho_c \sim 10^{0} \; {\rm nm}^3$.   Since for living cells $\rho^* \sim 10^{-3} \; \mathrm{nm}^{-3}$ \cite{cellbiobythe}, this suggests that random demixing due to fluctuations in volume is not likely to happen in living organisms.

\section*{References}
\bibliographystyle{ieeetr} 


\end{document}